\documentclass[amsmath, amssymb, aps, preprint, bibliography]{revtex4-1}
\usepackage{amssymb, amsmath, bm, graphicx,color}
\usepackage{epstopdf}
\usepackage{array}

\begin{document}


\title{Thermophotovoltaic energy conversion in far-to-near-field transition regime}


\author{Jaeman Song$^{1,2}$, Junho Jang$^{3}$, Mikyung Lim$^{4}$, Minwoo Choi$^{1,2}$, Jungchul Lee$^{1,2*}$, Bong Jae Lee$^{1,2}$}
\email{jungchullee@kaist.ac.kr, bongjae.lee@kaist.ac.kr\\}
\affiliation{1. Department of Mechanical Engineering, Korea Advanced Institute of Science and Technology, Daejeon 34141, South Korea\\
2. Center for Extreme Thermal Physics and Manufacturing, Korea Advanced Institute of Science and Technology, Daejeon 34141, South Korea\\
3. School of Electrical Engineering, Korea Advanced Institute of Science and Technology, Daejeon 34141, South Korea\\
4. Nano-Convergence Mechanical Systems Research Division, Korea Institute of Machinery and Materials, Daejeon 34103, South Korea\\
*Corresponding Authors\\
}


\date{\today}

\begin{abstract}
Recent experimental studies on near-field thermophotovoltaic (TPV) energy conversion have mainly focused on enhancing performance via photon tunneling of evanescent waves. In the sub-micron gap, however, there exist peculiar phenomena caused by the interference of propagating waves, which is seldom observed due to the dramatic increase of the radiation by evanescent waves in full spectrum range. Here, we experimentally demonstrate the oscillatory nature of near-field TPV energy conversion in the far-to-near-field transition regime (250-2600 nm), where evanescent and propagating modes are comparable due to the selective spectral response by the PV cell. Noticeably, it was possible to produce the same amount of photocurrent at different vacuum gaps of 870 and 322 nm, which is 10$\%$ larger than the far-field value. Considering the great challenges in maintaining nanoscale vacuum gap in practical devices, this study suggests an alternative approach to the design of a TPV system that will outperform conventional far-field counterparts.
\end{abstract}

\maketitle



A thermophotovoltaic (TPV) is an energy conversion device that directly converts photon energy into electrical energy; device consists of a high-temperature emitter and photovoltaic (PV) cell (i.e., receiver) \cite{datas2021thermophotovoltaic}. Based on a theoretical concept proposed in the early 2000s \cite{whale2002modeling, narayanaswamy2003surface}, when the gap between the emitter and the receiver becomes closer than the characteristic wavelength of thermal radiation determined by Wien’s displacement law, the near-field radiation associated with evanescent waves can contribute to photocurrent generation by the PV cell. Therefore, the energy conversion performance of a TPV device operating in the near-field regime can be improved in general, and such device is called a near-field thermophotovoltaic (NF-TPV).

Accordingly, most studies on NF-TPV experiments \cite{dimatteo2001enhanced, fiorino2018nanogap, inoue2019one, bhatt2020integrated, lucchesi2021near, inoue2021integrated, mittapally2021near} have taken extensive measures to reduce the gap between the emitter and the PV cell in a vacuum because the radiative heat flux by evanescent waves increases as the gap decreases. However, unlike the near-field radiative heat transfer, which takes advantage of the full spectral range, the PV cell can generate electrical energy exclusively from photons, whose energy is greater than the bandgap. Due to such a spectral difference, the performance of the NF-TPV does not necessarily increase at a smaller vacuum gap. When the vacuum gap becomes comparable to the characteristic wavelength of thermal radiation or narrower, one can observe distinctive thermal radiation originating not only from the evanescent mode (i.e., near-field effect) but also from the propagating mode (i.e., interference effect) \cite{narayanaswamy2016minimum, tsurimaki2017coherent}. The interference effect begins to emerge when the evanescent mode is spectrally suppressed by the PV cell, causing performance fluctuations of the TPV system, as opposed to an overwhelming near-field effect shading the interference effect in the general full spectrum regime. Such phenomena have already been addressed in several theoretical works \cite{park2008performance, song2019analysis, vaillon2019micron} after the first report by Whale \cite{whale2001influence}. Although power output fluctuations with respect to the vacuum gap were observed at the vacuum gap near 1 $\mu$m between the propagating-mode-dominant far-field regime and evanescent-mode-dominant near-field regime to some extent in refs.\ \cite{fiorino2018nanogap, mittapally2021near}, those researchers only emphasized the performance improvement due to the near-field effect. In other words, surprisingly, no NF-TPV experiment has yet paid attention to performance variation in the far-to-near-field transition regime.

In this work, we aim to experimentally demonstrate TPV energy conversion resulting from both propagating and evanescent modes in the far-to-near-field transition regime and to quantitatively evaluate the corresponding photocurrent generation. Using a 786-K-temperature doped-Si emitter and a Au/$n$-GaSb Schottky PV cell, the photocurrent and electrical power output in the PV cell are measured with the vacuum gap ranging from 250 nm to 2600 nm. Measurements are thoroughly compared with theoretical predictions, evaluating the independent contribution of evanescent and propagating waves. Taking one step further, a method to probe the spectral radiative heat flux across the nanoscale vacuum gap is proposed for the first time.

\begin{figure}[!b]
\centering\includegraphics[width=0.91\textwidth]{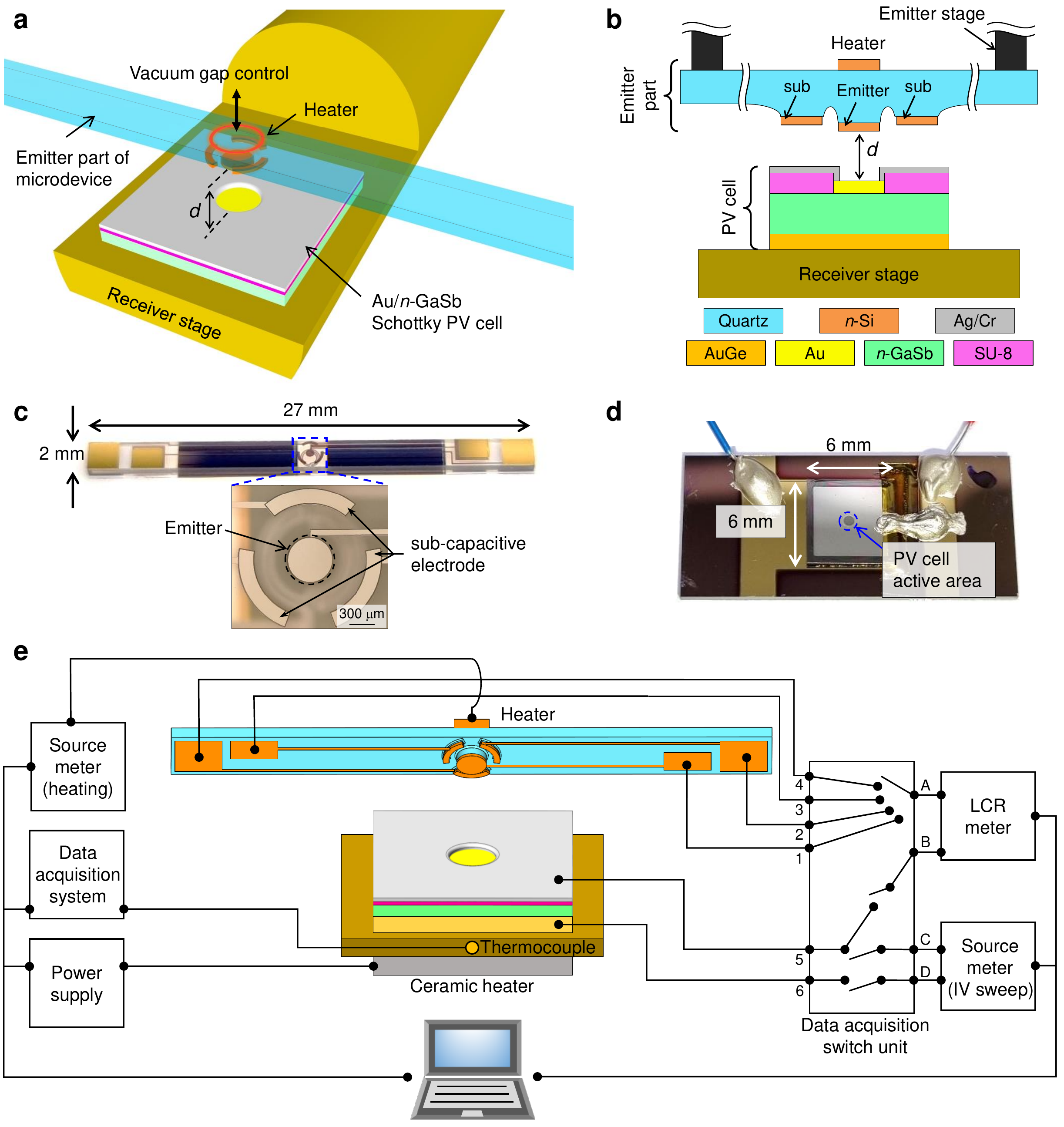}
\caption{\textbf{Configuration of near-field thermophotovoltaic (NF-TPV) system. a} Schematic illustration consisting of emitter part of microdevice and Au/$n$-GaSb Schottky PV cell. \textbf{b} Cross-sectional schematic of system. Materials constituting the emitter part of microdevice and PV cell are denoted. \textbf{c, d} Photographs showing emitter part of microdevice and Schottky-junction-based PV cell, respectively. The width and length of the emitter part of the microdevice are 27 mm and 2 mm, respectively. Microdevice is mounted on an emitter stage of a nanopositioner. The size of the PV cell is 6$\times$6 mm$^2$. PV cell is mounted on a receiver stage following attachment to chip carrier. \textbf{e} Schematic illustration of experimental setup.}
\label{Fig1}
\end{figure}

\section*{Results}
\subsection*{Devices and experimental setup}
To thoroughly investigate the performance of the TPV system in the far-to-near-field transition regime, we construct a vacuum gap tunable TPV by employing a microfabricated emitter and Schottky-junction-based PV cell pair, along with a precise nanopositioner. Figure \ref{Fig1}a provides a schematic illustration of the experimental setup. The emitter device is fabricated on a quartz wafer and diced to have a size of $27\times2$ mm$^2$. The 505.7-$\mu$m-diameter and 800-nm-thick $n$-doped-Si emitter faces the active area of the PV cell. A thin-film heater made of the same material as the emitter is located on the opposite side to raise the emitter temperature. The emitter part of the microdevice is firmly attached to the emitter stage of the nanopositioner such that the vacuum gap distance between the emitter and the PV cell can be precisely controlled with the resolution of $<15$ nm \cite{lim2018tailoring}. For the PV cell, a Au/$n$-GaSb Schottky-junction-based photodiode having a bandgap of 0.726 eV is fabricated \cite{jang2021analysis}. The overall size of the PV cell is $6\times6$ mm$^2$ and the diameter of the active area subject to the thermal radiation is 705.7 $\mu$m. Because the diameter of the active area of the PV cell is 200 $\mu$m larger than that of the emitter, any lateral misalignment of the emitter on the PV cell can be mitigated. After the PV cell is attached to the chip carrier, it is mounted on the oxygen-free copper stage (i.e., receiver stage in Figs.\ \ref{Fig1}a and b). Temperature of the PV cell can be controlled using a ceramic heater attached at the bottom of the receiver stage. The detailed fabrication process of the emitter and the PV cell are provided in Supplementary Note 1. As denoted in Fig.\ \ref{Fig1}b, an SU-8 passivation layer is patterned around the Au/$n$-GaSb Schottky junction of the PV cell and this layer is covered by a Ag/Cr layer, used as the upper electrode of the PV cell. On the other side, a AuGe layer acting as the lower electrode makes ohmic contact with the $n$-GaSb semiconductor.

Because the doped-Si emitter and the thin Au layer of the PV cell are designed to serve as capacitive electrodes in parallel, we can estimate the main vacuum gap distance (i.e., described as $d$ in Fig.\ \ref{Fig1}b) between the emitter and the PV cell by measuring the capacitance \cite{ottens2011near, lim2018tailoring, ying2019super, lim2020surface}. In addition, three doped-Si sub-capacitive electrodes (sub1, sub2, and sub3) are symmetrically fabricated around the emitter to obtain three additional vacuum gap distances between each sub-capacitive electrode and the Ag/Cr electrode of the PV cell. Those gaps are used to quantify and control the degree of parallelism between the emitter and the active area of the PV cell. Sub-capacitive electrodes are fabricated to be 2 $\mu$m lower than the emitter to prevent physical contact, which can be caused by the thickness of the SU-8 passivation layer. We can achieve parallelism to a reasonable level (i.e., tilt angle is $<8\times10^{-5}$ rad. To be discussed in PV cell characterization and vacuum gap control section) by controlling the emitter position through the nanopositioner and detecting parallelism via the three sub-capacitive electrodes. The specific configuration of the nanopositioner is provided in Supplementary Note 2.

\subsection*{Experiment procedure}
The experimental process is divided into two steps: the preliminary step and the major step. In the preliminary step, the temperature of the PV cell is kept constant, and the position of the emitter is controlled to be parallel with the PV cell. In the major step, the vacuum gap distance between the emitter and the PV cell is measured, followed by collecting a current-voltage ($I$-$V$) characteristic to measure the photocurrent and electrical power generated in the PV cell while the temperature of the emitter is raised by Joule heating the heater. A schematic of the experimental setup is provided in Fig.\ \ref{Fig1}e, including the emitter part of the microdevice, the PV cell, and the experimental instruments.

To keep the temperature of the PV cell constant in the preliminary step, the power supplied to the ceramic heater (CER-1-01-00540, Watlow) is feedback controlled such that the temperature, measured by the thermocouple inserted in the receiver stage, is maintained at 303 K. To check the parallelism between the emitter and the active area of the PV cell, the main and sub vacuum gap distances are sequentially measured using an LCR meter (E4980AL, Keysight) together with a data logger switch unit (34970A, Keysight). For example, to measure the main vacuum gap distance, we connect the data logger switch unit and the LCR meter through circuits 1-A and 5-B in Fig.\ \ref{Fig1}e. As circuit 1-A is switched to 2-A, 3-A, and 4-A, sub vacuum gap distances are sequentially acquired. Based on the data of sub vacuum gap distances, the parallelism can be adjusted using the nanopositioner attached to the emitter part of the microdevice.

In the major step, the voltage applied to the heater is feedback controlled using a sourcemeter (2400, Keithley) to supply constant electrical power. When the emitter temperature reaches the target temperature determined by the ANSYS transient thermal analysis on the basis of the input heating power data (see Supplementary Note 3), the main vacuum gap distance is measured by configuring the circuit of the switch unit to 1-A and 5-B. Then, the $I$-$V$ characteristic of the PV cell is obtained using another sourcemeter (2400, Keithley) by switching the circuit to 5-C and 6-D. Since the experimental setup is placed in a vacuum chamber ($<1\times10^{-3}$ Pa), the conduction and convection heat transfer by air are negligible. The configuration of the emitter part of the microdevice and the PV cell mounted on the nanopositioner is described in Supplementary Note 2.

\subsection*{PV cell characterization and vacuum gap control}
\begin{figure}[!b]
\centering\includegraphics[width=0.73\textwidth]{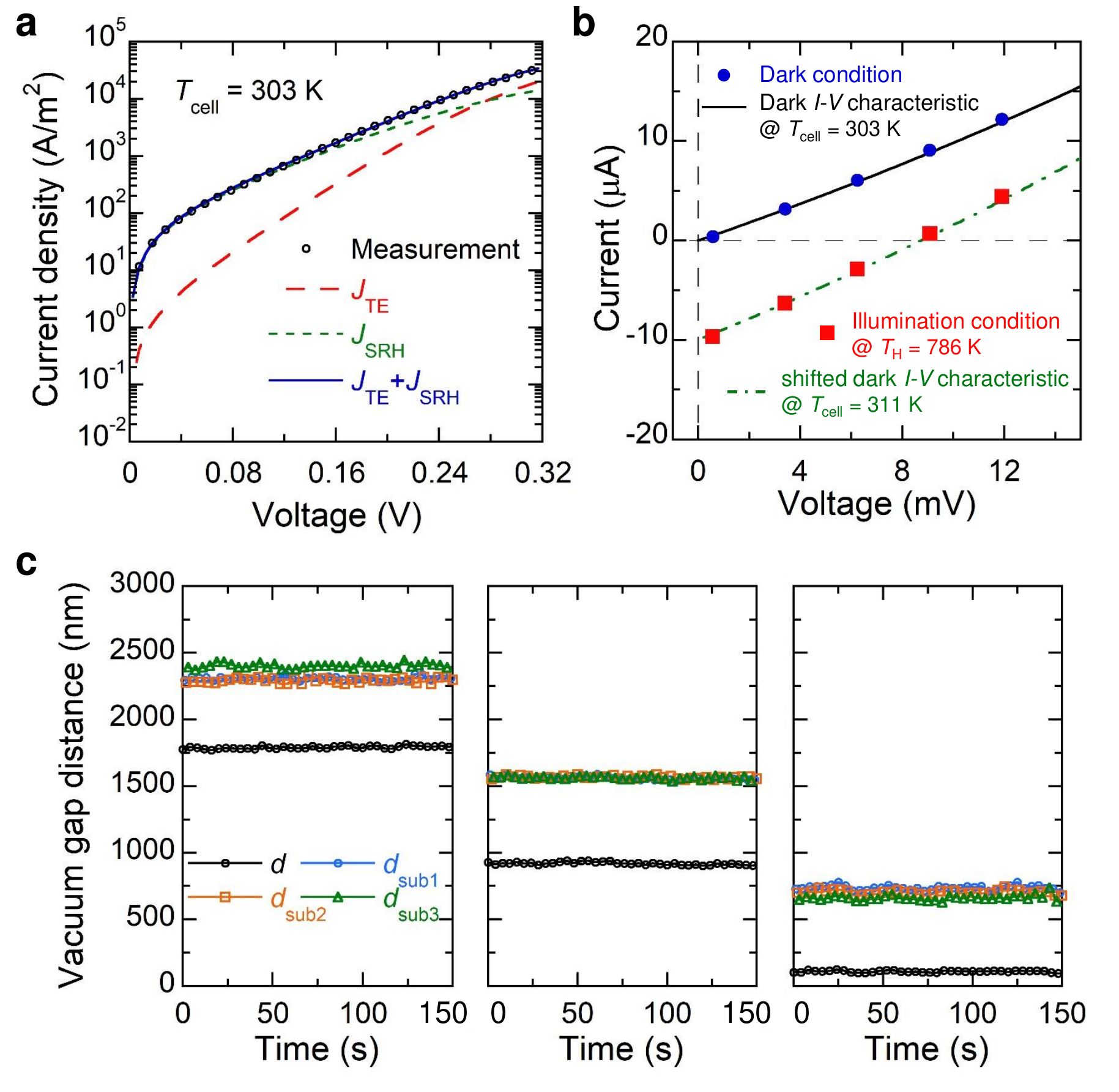}
\caption{\textbf{Characterization of experimental setup. a} Current density and voltage ($J$-$V$) characteristic of Au/$n$-GaSb Schottky PV cell at 303 K. Fitting results show that the sum of thermionic emission $J_\text{TE}$ and Shockley-Read-Hall (SRH) recombination $J_\text{SRH}$ current density describes the current mechanism of the Au/$n$-GaSb Schottky PV cell. \textbf{b} Measured $I$-$V$ values under dark (solid circles) and illumination conditions (solid squares). For the illumination condition, the temperature of the emitter is 786 K, and the vacuum gap distance is 2029 nm. By comparing measured values with $I$-$V$ curves acquired at various temperatures of the PV cell, we can verify that the PV cell is at a temperature of 311 K under the illumination condition. \textbf{c} Aligning the emitter and the PV cell parallel at three vacuum gap distances of 1791 nm, 919 nm, and 109 nm. The standard deviations are 10 nm, 10 nm, and 8 nm, respectively, for 150 seconds. Since the LCR meter is set to accurately measure the vacuum gap distance between the emitter and the PV cell, the absolute values of $d_\text{sub1-3}$ are not guaranteed. Thus, those values are used only to check the parallelism.}
\label{Fig2}
\end{figure}

To characterize the fabricated Au/$n$-GaSb Schottky PV cell, $I$-$V$ curves are obtained and analyzed according to the cell temperature using the multi-current fitting method \cite{jang2021analysis}. The total current flowing through the Schottky diode results from the sum of multiple independent current mechanisms such as thermionic emission current and recombination current \cite{donoval1991analysis}. In our previous report on a Au/$n$-GaSb Schottky diode, we demonstrated that the forward bias current of the diode is mainly composed of thermionic emission current density $J_\text{TE}$ and Shockley-Read-Hall recombination current density $J_\text{SRH}$ at temperatures above 240 K, while tunneling current, leakage current, and other recombination currents are negligible \cite{jang2021analysis}. Based on that report, Fig.\ \ref{Fig2}a shows that the fitting results expressed with $J_\text{TE}$ and $J_\text{SRH}$ agree well with the measured current density and voltage ($J$-$V$) curve for the Au/$n$-GaSb Schottky PV cell used in this work (see Supplementary Note 4 for detailed expression of $J$-$V$ characteristics). Figure \ref{Fig2}b shows the measured $I$-$V$ values at dark condition (blue circles) and at illumination condition (red squares) with the emitter temperature maintained at 786 K when the vacuum gap distance $d$ is 2029 nm. By absorbing the thermal radiation from the emitter, $I$-$V$ values under the illumination condition is shifted to the negative $y$-direction. In addition, the gradient of the $I$-$V$ values under the illumination condition is slightly steeper than that under the dark condition. This is because the radiation from the high-temperature emitter can also increase the temperature of the Schottky junction, which in turn changes the dark current of the Schottky diode. In other words, the junction temperature of the PV cell can be estimated from \textit{a priori} knowledge of the temperature-dependent $I$-$V$ curve. Using a ceramic heater attached to the bottom of the receiver stage, temperature-dependent $I$-$V$ curves are measured in advance in the temperature range of 296--333 K (see Supplementary Note 4 for details). Based on fitting analysis, the temperature of the PV cell is estimated to be increased to 311 K under the illumination condition (green dash-single dotted line in Fig.\ \ref{Fig2}b). The analyzed temperature-dependent $J$-$V$ characteristics can be regarded as the dark current of the PV cell.

Figure \ref{Fig2}c shows exemplary measurements of the main and sub vacuum gap distances for 150 seconds at three averaged main vacuum gap distances (i.e., 1791, 919, and 109 nm); the standard deviation is less than 10 nm. When we conservatively assume that the emitter is tilted relative to the PV cell as much as the maximum standard deviation of the sub vacuum gaps, the largest gap deviation within the emitter can be estimated to be about 40 nm, which is sufficiently small to be considered as being parallel within a 505.7-$\mu$m-diameter emitter area, i.e., tilt angle is $<8\times10^{-5}$ rad.

\begin{figure}[!b]
\centering\includegraphics[width=0.95\textwidth]{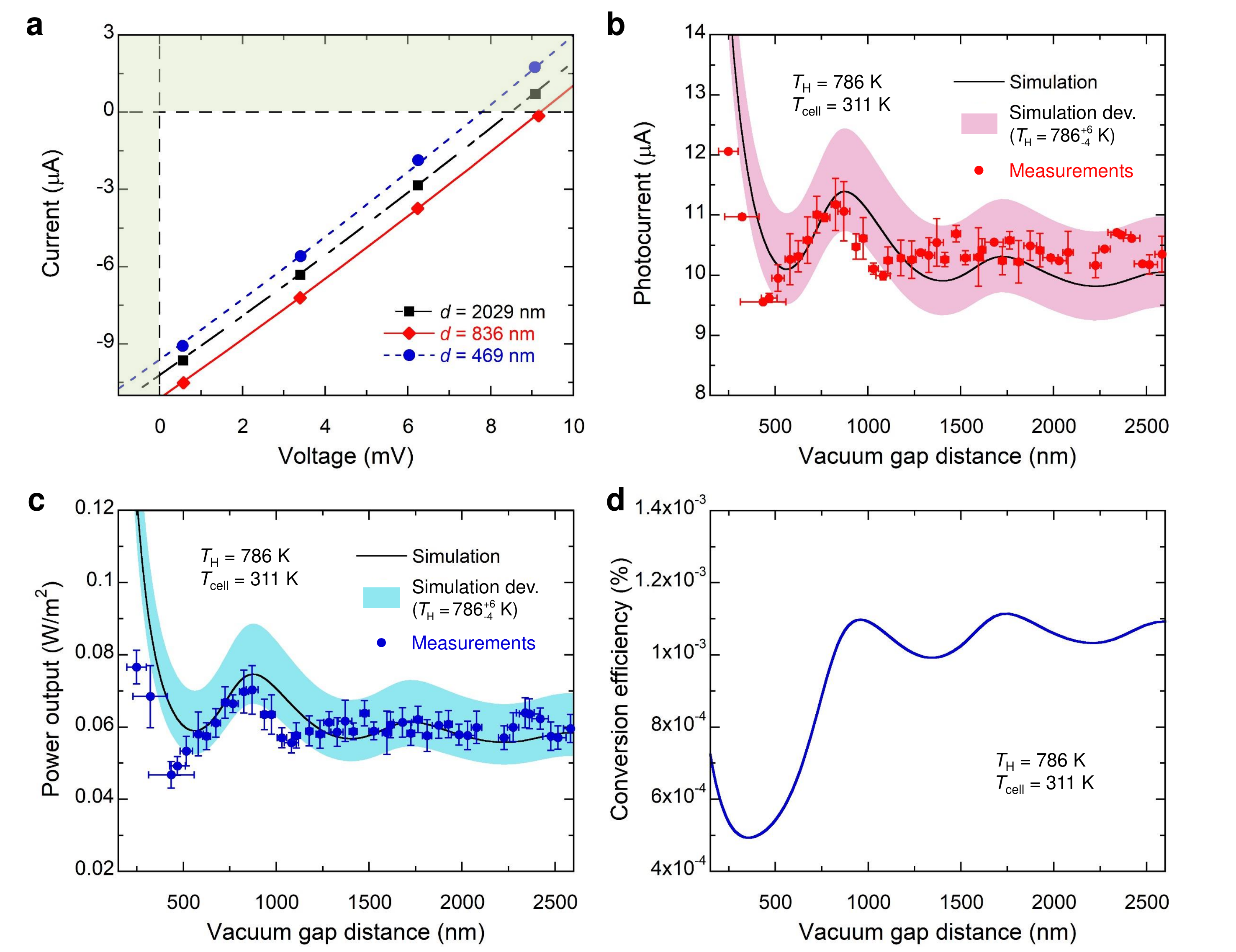}
\caption{\textbf{Performance of thermophotovoltaic system as function of vacuum gap distance when emitter temperature is 786 K. a} Current-voltage ($I$-$V$) characteristics at three vacuum gap distances. \textbf{b} Simulated and measured photocurrents. The colored band represents the photocurrent deviation, mainly attributed to the heating time deviation and the volume deviation of the graphite adhesive (see Supplementary Note 5). The $x$-axis and $y$-axis error bars represent measurement uncertainties of the vacuum gap distance and the photocurrent, respectively (see Supplementary Note 5). \textbf{c} Simulated and measured electrical power outputs. The colored band represents the electrical power deviation and the $x$-axis and $y$-axis error bars represent measurement uncertainties of the vacuum gap distance and the power output, respectively. \textbf{d} Simulated conversion efficiency.}
\label{Fig3}
\end{figure}

\subsection*{Experimental results}
The photocurrent generation and the electrical power output can be estimated by measuring the $I$-$V$ characteristic when the PV cell is exposed to a high-temperature emitter. Figure \ref{Fig3}a compares the three $I$-$V$ characteristics measured at three vacuum gap distances when the emitter is at 786 K. The short-circuit current of the $I$-$V$ curve (i.e., current corresponding to the zero-bias condition) can be regarded as photocurrent generation in the PV cell. The photocurrent measured at $d=836$ nm is 1.1 times greater than that measured at $d=2029$ nm. Counter-intuitively, the resulting photocurrent generation at $d=469$ nm is smaller than that at $d=2029$ nm. Since the slope of the $I$-$V$ curve remains nearly unchanged for the three vacuum gap distances, the PV cell's junction temperature can be considered as almost constant at 311 K, even though the total radiative heat flux changes according to the vacuum gap distance. Therefore, the PV cell itself is not responsible for the observation.

To elucidate this phenomenon, the photocurrent generation is measured as a function of the vacuum gap distance and compared with theoretical predictions based on the fluctuational electrodynamics and quantum efficiency of the PV cell (see Methods section). Figure \ref{Fig3}b reveals the oscillatory nature of the photocurrent with respect to the vacuum gap distance, unlike the monotonic increase of the heat transfer in the near field \cite{zhang2007nano, basu2016near}. This oscillation occurs because the PV cell utilizes photons whose energy is greater than the bandgap energy; that is, the PV cell acts as an optical short-pass filter. Due to the coherence of thermal radiation in the far-to-near-field transition regime, we provide the first experimental demonstration that the performance of the NF-TPV system is not always improved as the emitter closely approaches the PV cell. For instance, the photocurrent generation at $d=870$ nm is almost the same as that at $d=322$ nm. Considering the trade-off between the vacuum gap distance and the active area of the device, it would be advantageous to operate the system at $d= 870$ nm rather than at $d=322$ nm.

Figure \ref{Fig3}c shows the measured and simulated electrical power outputs produced by the PV cell as a function of the vacuum gap distance. Electrical power output is determined by the maximum product of $I$ and $V$ on the acquired $I$-$V$ curve. The theoretical $I$-$V$ characteristics under illumination condition are determined as the difference between the photocurrent and the dark current (see Methods section). Similar to the case of the photocurrent, an oscillatory nature can also be observed in the measured electrical power output. At the smallest vacuum gap of 248 nm, we were able to achieve power output of 0.077 W/m$^2$, which is 1.3 times greater than that at the 2584 nm gap. The corresponding conversion efficiency of the NF-TPV, which can be calculated by dividing the calculated electrical power output by the radiative heat flux, is shown in Fig.\ \ref{Fig3}d. For $d>1000$ nm, the conversion efficiency follows a trend similar to that of the photocurrent generation or electrical power output. For $350\text{ nm}<d<1000$ nm, however, the conversion efficiency decreases dramatically as the gap decreases because the increase in radiative heat flux by evanescent mode is much larger in the sub-bandgap spectral region than in the above-bandgap spectral region. Therefore, for the considered Schottky-junction-based PV cell, operation in the near-field regime is not necessary unless one can make a practical device at sub-100-nm vacuum gap.

\section*{Discussion}
\begin{figure*}[!b]
\centering\includegraphics[width=0.95\textwidth]{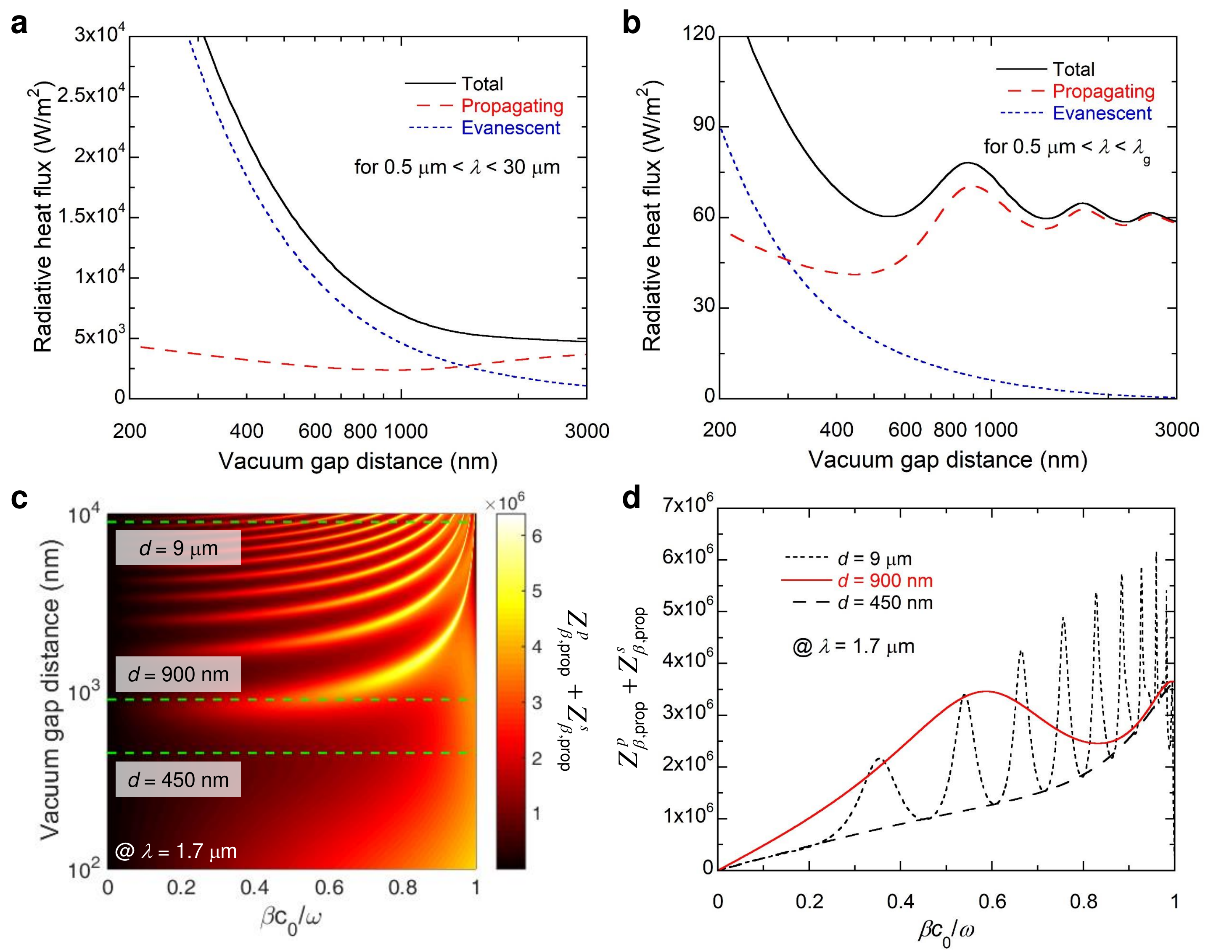}
\caption{\textbf{Physical mechanism of oscillatory nature of thermophotovoltaic system.} Total, propagating, and evanescent components of radiative heat flux between doped-Si emitter and Au/$n$-GaSb PV cell in wavelength ranges: \textbf{a} from 0.5 $\mu$m to 30 $\mu$m and \textbf{b} from 0.5 $\mu$m to $\lambda_g$ (i.e., 1.72 $\mu$m). \textbf{c} Contour plot of propagating mode of exchange function in gap size and normalized parallel wavevector space at wavelength of 1.7 $\mu$m. \textbf{d} Propagating mode of exchange function according to normalized parallel wavevector for three vacuum gap distances of 9 $\mu$m, 900 nm, and 450 nm at wavelength of 1.7 $\mu$m.}
\label{Fig4}
\end{figure*}

To better explain the physical mechanism of the oscillatory nature of the TPV energy conversion with respect to the vacuum gap distance, we evaluate the individual contributions of evanescent and propagating waves to the radiative heat flux. In Fig.\ \ref{Fig4}a, the radiative heat flux between 800-nm-thick doped-Si-covered SiO$_2$ and the Au/$n$-GaSb Schottky PV cell is plotted in the wavelength range from 0.5 $\mu$m to 30 $\mu$m. In general, the evanescent waves (i.e., frustrated modes) start to contribute to the radiative heat flux when the vacuum gap distance is comparable to the characteristic wavelength of thermal radiation at the emitter temperature. Although the contribution of the propagating mode shows a minimum around $d=1000$ nm, the total radiative heat flux does not exhibit such a valley due to the evident contribution of frustrated modes at this gap. However, the situation changes if one considers only the photons with energy above that of the bandgap. The radiative heat flux is now calculated only in the wavelength range from 0.5 $\mu$m to the bandgap wavelength of GaSb (i.e., 1.72 $\mu$m), as shown in Fig.\ \ref{Fig4}b. With the PV cell, which acts as a short-pass filter, the contribution of frustrated modes is greatly suppressed near $d=1000$ nm. This is because photon tunneling accompanied by evanescent waves with shorter wavelengths occurs more favorably at smaller vacuum gaps. For the same reason, the vacuum gap distance at which contributions by frustrated and propagating modes are equal to each other, as can be seen in Fig.\ \ref{Fig4}a (i.e., $d\sim$1450 nm), shifts towards much smaller values in Fig.\ \ref{Fig4}b (i.e., $d\sim$300 nm). With the use of the PV cell, the interference fringes of propagating waves are greatly amplified. Notice that the above-bandgap radiative heat flux in Fig.\ \ref{Fig4}b resembles the measurements in Figs.\ \ref{Fig3}b and c. Although the same photocurrents are generated at $d=870$ nm and $d=322$ nm, the above-bandgap radiation mechanisms contributing to these are different. At $d=870$ nm, the contribution of propagating modes is 90.0$\%$, but this drops to 53\% at $d=322$ nm. Due to the increased contribution of frustrated modes at $d=322$ nm, the sub-bandgap absorption also increases via near-field effects, remarkably reducing the conversion efficiency (see Fig.\ \ref{Fig3}d).

The physics underlying the interference effect can be revealed in a contour plot of the propagating exchange function (see Methods section) as a function of the normalized parallel wavevector component and the vacuum gap distance, as shown in Fig.\ \ref{Fig4}c. Since the temperature of the emitter is not high enough compared to the bandgap of the PV cell, we calculate the exchange function at the wavelength of 1.7 $\mu$m, given that the dominant portion of the radiation is transferred near the bandgap wavelength. Figure \ref{Fig4}c clearly shows the interference fringes in the gap-wavevector space. When the vacuum gap is relatively large, the propagating component of the radiative heat flux is transferred by multiple fringes; the number of fringes contributing to the heat transfer decreases as the vacuum gap distance gets smaller. As a representative illustration, the exchange function of the far-field (i.e., at $d=9$ $\mu$m), maximum (i.e., at $d=900$ nm), and minimum (i.e., at $d=450$ nm) cases are compared in Fig.\ \ref{Fig4}d. Since multiple peaks of the exchange function can be employed in the far-field case, almost constant radiation can be delivered regardless of the number or position of peaks, which can change depending on the vacuum gap distance (i.e., incoherent radiation). For the maximum case (i.e., at $d=900$ nm), the propagating radiative heat flux increases because the broadly enhanced exchange function exploits the interference branch that is widely formed along the parallel wavevector. On the other hand, when $d=450$ nm, the propagating radiative heat flux is less than the far-field value because almost no interference branch participates in the radiation. The vacuum gap where the peak photocurrent appears due to interference can be readily predicted using the intensity maximum condition for the Fabry-Perot interferometer (i.e., $d_m=m\lambda/2$, where $m$ is the order of the Fabry-Perot interference). At $\lambda=\lambda_g=1.72$ $\mu$m, $d_m= 860$ nm, 1720 nm, 2580 nm, $\dotsb$, and these values agree reasonably well with the vacuum gaps at which photocurrent or power output peaks are located in Figs.\ \ref{Fig3}b and c. Furthermore, the quantitative agreement between the measurements and predictions in Fig.\ \ref{Fig3} also suggests that the fluctuational electrodynamics is capable of capturing the physics in the far-to-near-field transition regime as well. 

Since the concept of the NF-TPV system was first proposed in ref.\ \cite{whale2002modeling}, there have been only a few groups who have experimentally realized NF-TPVs \cite{dimatteo2001enhanced, fiorino2018nanogap, inoue2019one, bhatt2020integrated, lucchesi2021near, inoue2021integrated, mittapally2021near}, despite the substantial progress in measuring the near-field radiation \cite{ottens2011near, lim2018tailoring, fiorino2018giant, ghashami2018precision, ying2019super, lim2020surface, desutter2019near, tang2020near}. The main factors of experimental difficulty are the essential requirements for a practical high-power-output NF-TPV: a large heat transfer area and a large temperature difference between the emitter and the PV cell. These conditions make maintaining the nanoscale vacuum gap quite challenging because of structural deformation by thermal stress \cite{ghashami2020experimental}. Recently, it was reported that a one-chip NF-TPV device with an area of 1 mm$^2$, a vacuum gap $<$140 nm, and a temperature difference of $\sim$900 K had been fabricated \cite{inoue2021integrated}. It was discussed that, to improve system efficiency, an up-scaled device is required to further reduce the conduction loss and thermal radiation loss. When the area becomes larger, it would be much more challenging to manufacture a device without any physical contact between the emitter and the PV cell for such a small vacuum gap. Therefore, exploiting the coherence of thermal radiation to improve the power output and conversion efficiency at experimentally preferred vacuum gap distances could be a more effective and practical approach for realizing highly efficient NF-TPV devices. 

Because the emitter temperature of 786 K is relatively low compared to the GaSb bandgap energy, the spectral peak of thermal radiation is located at a wavelength much larger than the bandgap wavelength. Therefore, if the photocurrent is calculated by multiplying the spectral photon flux by the internal quantum efficiency and the electron volt, the spectral shape seems as if the light passed through a 1.72-$\mu$m bandpass filter (see Supplementary Note 6). That is, the inherent Planck distribution of the emitter acts as a long-pass filter for the wavelength, but the bandgap of the PV cell act as a short-pass filter. Consequently, it can be regarded that the present work measures a narrow-band radiative heat transfer with the full width at half maximum of approximately 180 nm. Although spectral measurement of near-field radiation has been considered difficult, the PV cell offers a direct way to measure narrow-band radiative heat flux. If the bandgap energy of the PV cell can be well controlled by changing the alloy composition in III-V ternary or quaternary compound semiconductors or the temperature of the PV cell, it will be possible to measure the spectral near-field radiation in the broad infrared region. 

We have proposed a systematic approach to analyze the performance of the TPV system in the far-to-near-field transition regime (250-2600 nm) by relying on a microfabricated emitter and Schottky-junction-based PV cell, along with a precise nanopositioner. We show that the PV cell behaves as a short-pass filter for spectral radiative heat flux and magnifies the contribution of propagating waves near the 1-$\mu$m vacuum gap, enabling measurement of the coherence of thermal radiation. The reason for the identical magnitude of photocurrent produced at different vacuum gaps of 870 and 322 nm is clearly proven to be the interference of propagating modes in the far-to-near-field transition regime. Generated photocurrent and electrical power agree well with theoretical predictions, confirming that fluctuational electrodynamics is capable of capturing the physics in the far-to-near-field transition regime. Considering the great challenges in maintaining the nanoscale vacuum gap in practical devices with increased surface areas, this study opens a new avenue for research by exploiting the coherence of thermal radiation to improve the power output and conversion efficiency at experimentally preferred vacuum gap distances. Furthermore, it is found that the use of PV cells can be extended to the measurement of spectral enhancement via surface polaritons which will be widely exploited to enhance the performance of the NF-TPV system.

\clearpage
\section*{Methods}
\subsection*{Near-field thermal radiation}
The near-field radiation between the emitter and the PV cell is calculated by adding contributions of propagating and evanescent modes, as \cite{zhang2007nano,basu2016near,lim2020surface}:
\begin{equation}
\begin{split}
q & = \int_0^{\infty} d\omega\ q_{\omega} =\int_0^{\infty} d\omega\ (q_{\text{prop},\omega} + q_{\text{evan},\omega}) \\
& = \sum_{j=p,s}\int_0^{\infty} d\omega\ \frac{\Theta(\omega, T_1)-\Theta(\omega, T_2)}{4\pi^2} \Big[\int_0^{\omega/c_0} Z^{j}_{\text{prop},\beta, \omega}(\beta, \omega) d\beta + \int_{\omega/c_0}^{\infty} Z^{j}_{\text{evan},\beta, \omega}(\beta, \omega) d\beta \Big]
\end{split}
\end{equation}
where $\Theta(\omega, T_i) = \hbar\omega/\{\text{exp}\left[\hbar\omega/(k_BT_i)\right]-1\}$ is the mean energy of the Planck oscillator, $\omega$ is the angular frequency, $\hbar$ is the reduced Planck constant, and $k_B$ is the Boltzmann constant. $T_1$ and $T_2$ are the temperature of the emitter and the PV cell, respectively. The exchange functions for propagating and evanescent waves are expressed by:
\begin{equation} \begin{aligned}
&Z_{\text{prop}, \beta, \omega}^{p,s}(\beta, \omega) = \frac{\beta (1-|r_{01}^{p,s}|^2)(1-|r_{02}^{p,s}|^2)}{|1-r_{01}^{p,s} r_{02}^{p,s} e^{i2k_{0z}d}|^2} \\ 
& Z_{\text{evan}, \beta, \omega}^{p,s}(\beta, \omega) =\frac{4\beta \text{Im}(r_{01}^{p,s})\text{Im}(r_{02}^{p,s}) e^{-2 \text{Im}(k_{0z})d}}{|1-r_{01}^{p,s} r_{02}^{p,s} e^{i2k_{0z}d} |^2},
\end{aligned} \end{equation}
where $r_{01}^{p,s}$ and $r_{02}^{p,s}$ are the modified reflection coefficients for the vacuum/doped-Si/SiO$_2$ (or vacuum/SiO$_2$) and vacuum/Au/$n$-GaSb multilayers, respectively. These values are obtained using Airy's formula \cite{biehs2007thermal,yeh1988optical}. $k_{0z}$ is the normal component of the wavevector in vacuum and Im() takes the imaginary part of a complex value. The dielectric function of SiO$_2$ is obtained from tabular data in \cite{palik1998handbook}. The thin-Au-film permittivity is obtained from the Drude model \cite{ordal1988optical}, including the electron-boundary scattering effect \cite{ijaz1978electron,ding2015thickness}. The frequency-, temperature-, and doping-concentration-dependent dielectric functions are used for $n$-doped Si and $n$-doped GaSb. For $n$-doped Si, the high-temperature-available dielectric function model \cite{lee2005temperature} is used. For $n$-doped GaSb in the sub-bandgap frequency regime, the Lorentz-Drude (LD) oscillator model is used to consider the absorption by lattice and free carriers \cite{patrini1997optical}. After we obtain the interband absorption coefficient in the above-bandgap frequency regime through the same process introduced in \cite{vaillon2019micron}, the real part of the refractive index is determined by the Kramers-Kr$\ddot{\text{o}}$nig relation. Supplementary Figure 12 shows the absorption coefficient of $n$-GaSb for a doping concentration of $5.3\times10^{16}$ cm$^{-3}$ at 303 K.


\subsection*{Simulation of photocurrent and electrical power output in the PV cell}

When photocurrent generation is simulated, we should consider the area difference between the emitter and the PV cell. The photocurrent generation is expressed by:
\begin{equation}
I_{ph}(d) = A_1e \int_0^{\infty}\frac{\eta_\text{int}(\omega)}{\hbar\omega}q_{\omega,1}(d)d\omega + (A_2 - A_1)e\int_0^{\infty}\frac{\eta_\text{int}(\omega)}{\hbar\omega}q_{\omega,2}d\omega
\end{equation}
where $A_1$ is area of emitter, $A_2$ is active area of PV cell, $\eta_\text{int}$ is frequency-dependent internal quantum efficiency (see Supplementary Note 7), $q_{\omega,1}$ is vacuum-gap-dependent spectral heat flux between doped-Si-covered SiO$_2$ and PV cell, and $q_{\omega,2}$ is spectral heat flux between SiO$_2$ and PV cell at 15-$\mu$m vacuum gap distance. The view factor between the doped-Si emitter and the $A_1$ area of the PV cell is larger than 0.99 at $d\lesssim$ 3000 nm, the range in which we are interested in this article. Such a high view factor guarantees that the radiation emitted from the doped-Si emitter reaches only the $A_1$ area of the PV cell; in turn, the remaining PV cell area $A_2-A_1$ receives the far-field radiation emitted by SiO$_2$. The $I$-$V$ characteristic of the PV cell under illumination is written as:
\begin{equation}
\begin{split}
I(V,d) & = I_{ph}(d) - I_\text{dark}(V) \\
& = I_{ph}(d) - A_3[J_\text{TE}(V) + J_\text{SRH}(V)]
\end{split}
\end{equation}
where $A_3$ is area of Au layer deposited on $n$-GaSb (i.e., 900-$\mu$m diameter. See Supplementary Note 1). Then, the maximum electrical power output density is calculated by $P_E(d)=\text{Max}[I(V,d)\cdot V]/A_2$. The conversion efficiency can be determined by dividing the electrical power by the near-field radiative heat flux absorbed in the PV cell [i.e., $\int_0^{\infty}(q_{\omega,1}+q_{\omega,2})d\omega$].

\section*{\NoCaseChange{Data availability}}
The data that support the findings of this study are available from the corresponding authors upon reasonable request.

\providecommand{\noopsort}[1]{}\providecommand{\singleletter}[1]{#1}%

\subsection*{Acknowledgments}
This research is supported by the Basic Science Research Program (NRF-2019R1A2C2003605 and NRF-2020R1A4A4078930) through the National Research Foundation of Korea (NRF) funded by Ministry of Science and ICT.

\subsection*{Author contributions}
J.S., M.L., J.L., and B.J.L. conceived the work. J.S. and J.J. designed and fabricated the MEMS-based microdevice and the PV cell. J.S. conducted the experiments and the calculations. J.J. performed the PV cell characterization. M.C. estimated the emitter temperature. All the authors contributed to data analysis. J.S., J.L., and B.J.L. wrote the paper with comments from all the authors.

\subsection*{Competing interests}
The authors declare no competing financial interests.

\end{document}


\setstretch{1.5}
\section*{Supplementary Information:\\
Thermophotovoltaic energy conversion in far-to-near-field transition regime}

\textbf{Authors}: Jaeman Song$^{1,2}$, Junho Jang$^{3}$, Mikyung Lim$^{4}$, Minwoo Choi$^{1,2}$, Jungchul Lee$^{1,2*}$, Bong Jae Lee$^{1,2*}$\\

\textit{1. Department of Mechanical Engineering, Korea Advanced Institute of Science and Technology, Daejeon 34141, South Korea\\
2. Center for Extreme Thermal Physics and Manufacturing, Korea Advanced Institute of Science and Technology, Daejeon 34141, South Korea\\
3. School of Electrical Engineering, Korea Advanced Institute of Science and Technology, Daejeon 34141, South Korea\\
4. Nano-Convergence Mechanical Systems Research Division, Korea Institute of Machinery and Materials, Daejeon 34103, South Korea\\
}\\
*Corresponding authors\\
*e-mail: jungchullee@kaist.ac.kr (Jungchul Lee), bongjae.lee@kaist.ac.kr (Bong Jae Lee)\\

\bigskip
\bigskip
\bigskip
\underline{\textbf{Table of Contents}}\\
1. Fabrication of emitter part of microdevice and Schottky PV cell\\
2. Positioner mounting emitter part of microdevice and PV cell\\
3. Emitter temperature estimation and validation\\
4. PV cell temperature estimation and current-voltage characterization\\
5. Data processing and uncertainty analysis\\
7. Quantum efficiency of Au/$n$-GaSb Schottky PV cell\\

\setstretch{1.5}

\clearpage
\section*{Supplementary Note 1. Fabrication of emitter part of microdevice and Schottky PV cell}

We fabricated two different MEMS devices for the sub-micron gap thermophotovoltaic (TPV) system experiment. One is an emitter part of a microdevice containing an emitter, three sub-capacitive electrodes, and a heater made of doped Si. The other is an Au/$n$-GaSb Schottky-junction-based PV cell.

\begin{figure}[h]
\centering\includegraphics[width=14cm]{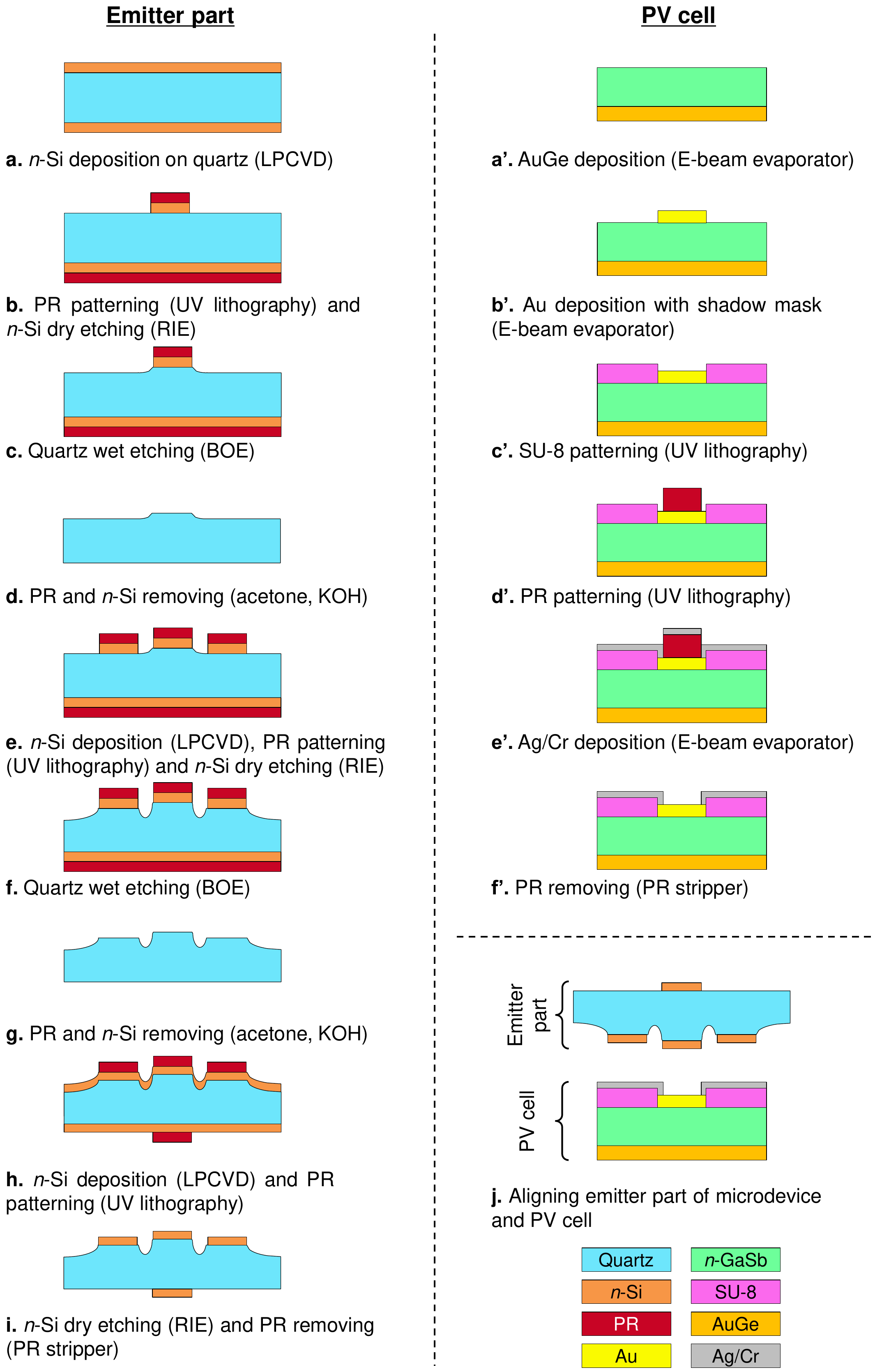}
\caption{\label{FigS1}Supplementary Figure 1: MEMS-fabrication process of microdevices. (a-i) Emitter part fabrication process. (a'-f') Schottky-junction-based PV cell fabrication process. (j) Aligned emitter part of microdevice and PV cell.}
\end{figure}

At the beginning of the fabrication of the emitter part of the microdevice, a quartz wafer is cleaned with Piranha solution (H$_2$SO$_4$:H$_2$O$_2$=1:2) for 15 minutes. After that, the 500-nm-thick doped Si is deposited using low pressure chemical vapor deposition (LPCVD) (Supplementary Fig.\ \ref{FigS1}a). The photoresist (PR, AZ9260) is spin coated on the surface and patterned using UV lithography. Then, doped Si is etched using reactive ion etching (RIE). Patterned doped Si is used as a mask for wet etching (Supplementary Fig.\ \ref{FigS1}b). We put the sample in buffered oxide etchant (6:1 BOE) for 25 minutes to wet etch the quartz; this creates a step difference between the emitter and three sub-capacitive electrodes (Supplementary Fig.\ \ref{FigS1}c). PR is removed using acetone and ethanol, following which the sample is immersed in 95$^{\circ}$C KOH to remove doped Si used as etch mask (Supplementary Fig.\ \ref{FigS1}d). In the same way, as shown in Supplementary Fig.\ \ref{FigS1}b, the deposition and lithography process is repeated with a different photomask (Supplementary Fig.\ \ref{FigS1}e). By wet etching the quartz surface with BOE for 150 minutes, a trench is formed in the area, excluding the emitter and sub-capacitive electrodes (Supplementary Fig.\ \ref{FigS1}f). This trench prevents physical contact between the emitter part and the PV cell and suppresses unnecessary vacuum-gap-dependent radiation. In the manner shown in Supplementary Fig.\ \ref{FigS1}d, PR and doped Si are removed (Supplementary Fig.\ \ref{FigS1}g). To make a 500-$\mu$m-diameter emitter and three sub-capacitive electrodes around it, the surface is patterned using PR and UV lithography, and 800-nm-thick doped Si is etched with RIE. On the backside, the heater is made in the same way. The remaining PR masks are removed using a stripper (ST-1023) (Supplementary Fig.\ \ref{FigS1}h-i).

To fabricate the PV cell, first, the native oxide layer of a $n$-GaSb wafer (Wafer Technology Ltd.) is removed using HCl solution (35-37$\%$ concentration) and the wafer is rinsed by isopropyl alcohol (IPA). Then, an ohmic contact is formed by annealing the sample in a furnace right after depositing a 150-nm-thick AuGe layer on the backside of the wafer using an e-beam evaporator (Supplementary Fig.\ \ref{FigS1}a'). This metal layer is used as a bottom electrode of the PV cell. On the upper surface, a 3.5-nm-thick 900-$\mu$m-diameter Au layer is deposited using an e-beam evaporator and a shadow mask (Supplementary Fig.\ \ref{FigS1}b'). Then, using SU-8 and UV lithography, an 800-$\mu$m-diameter hole-shaped PR passivation layer is fabricated (Supplementary Fig.\ \ref{FigS1}c'). We used SU-8 to prevent this passivation layer from being removed in a later lift-off process. For the lift-off process, PR (AZnlof 2035) is patterned with a diameter of 700 $\mu$m using UV lithography (Supplementary Fig.\ \ref{FigS1}d'). The size of this PR mask determines the size of the active area of the PV cell. As an upper electrode, an Ag/Cr: 120 nm/10 nm metal layer is deposited using an e-beam evaporator (Supplementary Fig.\ \ref{FigS1}e'). After that, the AZnlof 2035 PR layer is removed using a stripper (ST-1023). Supplementary Fig.\ \ref{FigS1}j shows the aligned configuration of the emitter part of the microdevice and the PV cell.

\begin{figure}[h]
\centering\includegraphics[width=17cm]{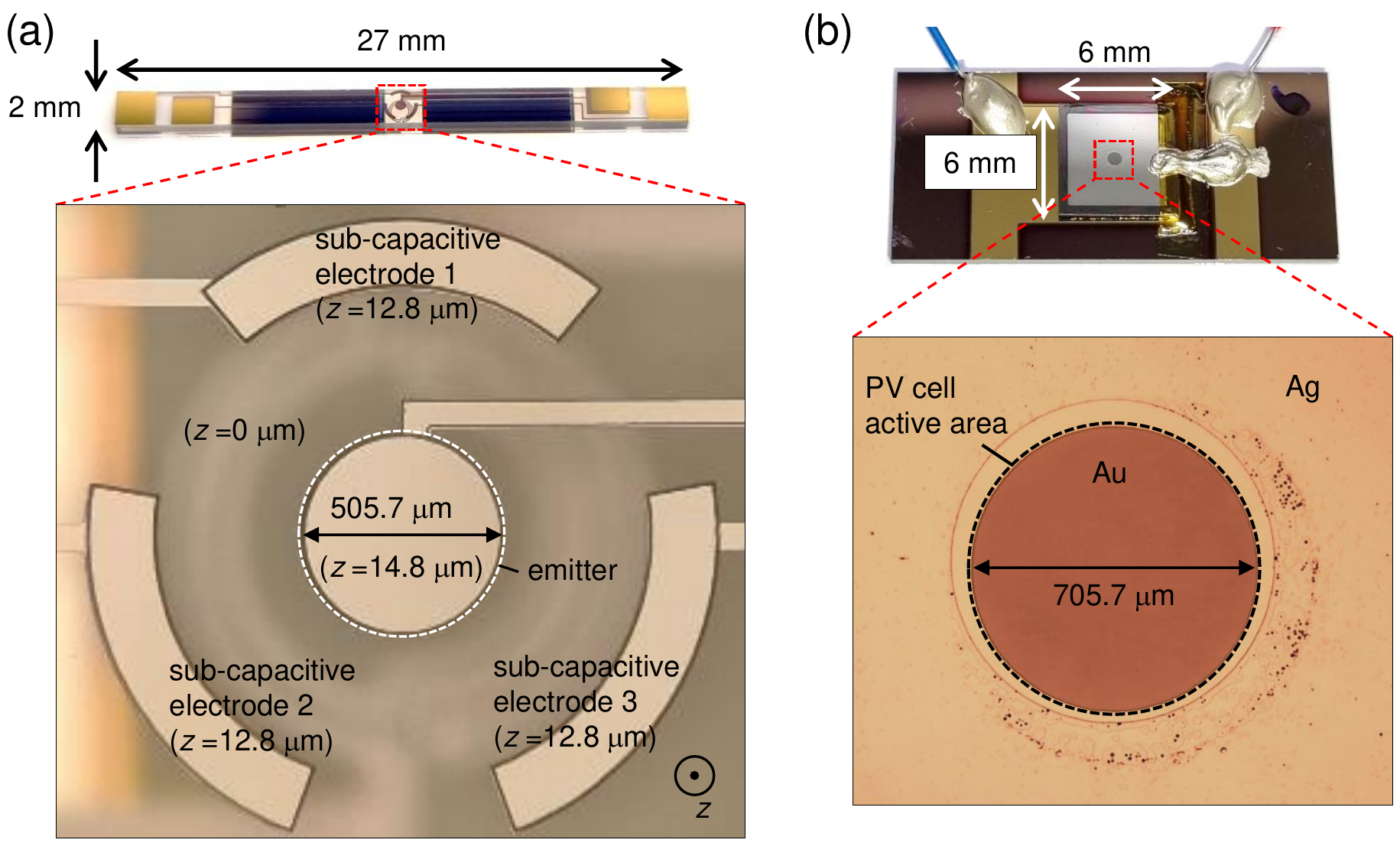}
\caption{\label{FigS2}Supplementary Figure 2: Images of fabricated devices. (a) Emitter part of microdevice. (b) Au/$n$-GaSb Schottky PV cell.}
\end{figure}

Supplementary Figs.\ \ref{FigS2}a and b provide images of the emitter part of the microdevice and the Au/$n$-GaSb Schottky PV cell, respectively. The size of the emitter part of the microdevice is $27\times2$ mm$^2$. Two gold-deposited capacitor soldering pads are placed at each end. In the enlarged image, we can see that the diameter of the emitter is 505.7 $\mu$m, and three sub-capacitive electrodes surround it. Using a stylus profilometer (AS500, KLA Tencor), the etching depth can be checked; it is found to be 2 $\mu$m for the sub-capacitive electrode and 14 $\mu$m for the trench. The height difference of the emitter part surface including the thickness of doped Si is denoted in Supplementary Fig.\ \ref{FigS2}a. Supplementary Fig.\ \ref{FigS2}b shows a $6\times6$ mm$^2$ Schottky PV cell attached on a chip carrier. The diameter of the active area is 705.7 $\mu$m.

\clearpage
\section*{Supplementary Note 2. Positioner mounting emitter part of microdevice and PV cell}

The emitter part of the microdevice and the Schottky PV cell are mounted to the positioner to control the vacuum gap distance between the emitter and the PV cell. In Supplementary Fig.\ \ref{FigS3}a, we can see the location of the PV cell and how the receiver stage is supported in the vacuum chamber. The emitter stage is placed on three motion guides through kinematic joints using zirconia balls (see Supplementary Fig.\  \ref{FigS3}b). Looking more specifically at the positioner in the design illustrations provided in Supplementary Fig.\ \ref{FigS4}, it is a combination of a 10-$\mu$m-moving resolution three-axis positioner (SLD125-LM-2, ST 1), a nanopositioner composed of three picomotor actuators (8302-V, Newport. $<30$ nm moving resolution) and three motion guides. We can align the emitter and the PV cell by adjusting the emitter's bulk $x$ and $y$ motions via the three-axis positioner. After that, the nanoscale $z$, roll, and pitch motions of the emitter are controlled by delivering the movement of the picomotor actuators to the emitter stage through the motion guides. As shown in Supplementary Fig.\ \ref{FigS5}a, the emitter part of the microdevice and the PV cell are aligned without physically hindering each other. The emitter part of the microdevice is mechanically fastened using two resin-based fixtures (see Supplementary Fig.\ \ref{FigS5}b). In Supplementary Fig.\ \ref{FigS5}c, it can be seen that the PV cell attached to the chip carrier is fixed on the receiver stage. To control the temperature of the PV cell, a thermocouple is inserted into the receiver stage and a ceramic heater is attached at the bottom of the receiver stage.

\begin{figure}[h]
\centering\includegraphics[width=17cm]{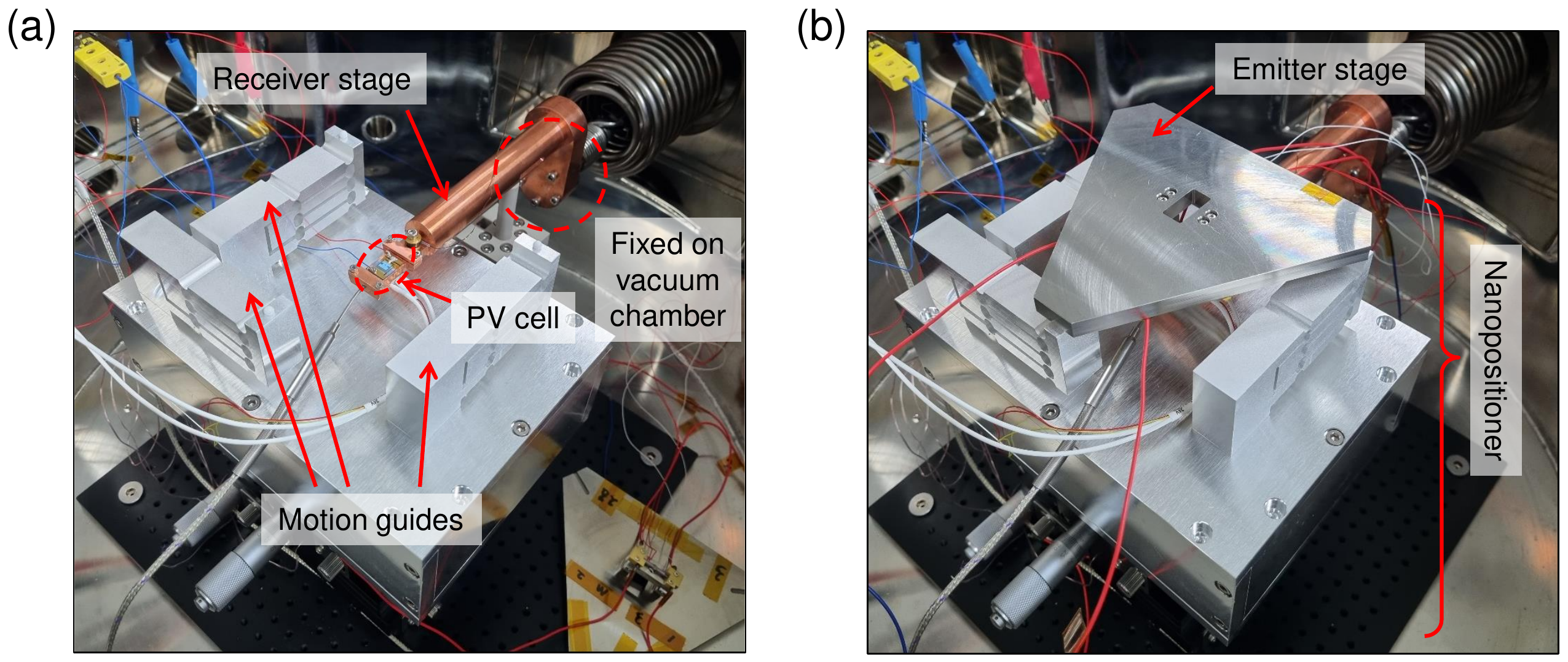}
\caption{\label{FigS3}Supplementary Figure 3: (a) Real image of positioner without emitter stage. (b) Real image of full positioner.}
\end{figure}

\begin{figure}[h]
\centering\includegraphics[width=17cm]{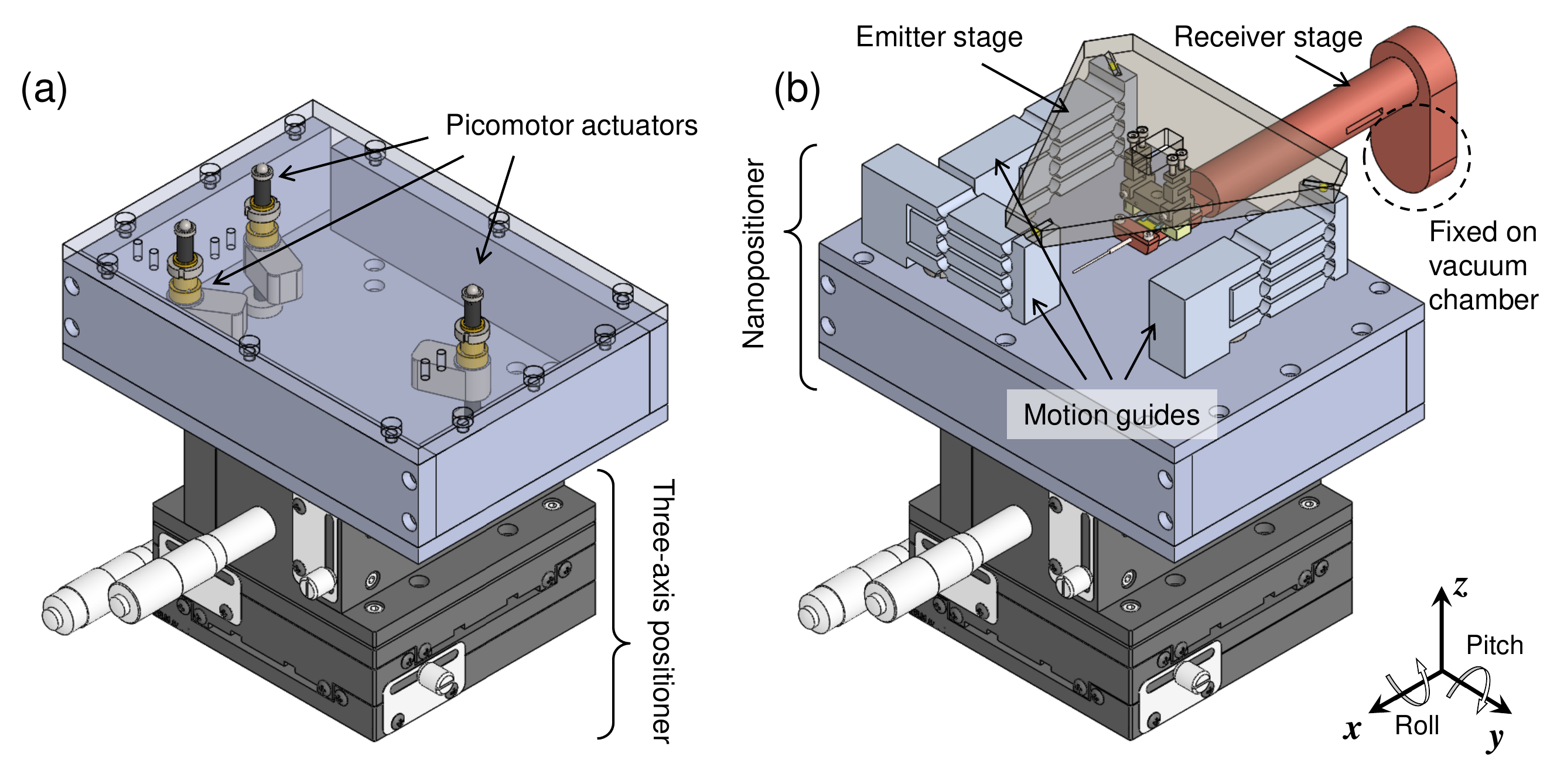}
\caption{\label{FigS4}Supplementary Figure 4: (a) Positioner design illustration excluding three motion guides, emitter stage, and receiver stage. (b) Full design illustration of positioner.}
\end{figure}

\begin{figure}[h]
\centering\includegraphics[width=17cm]{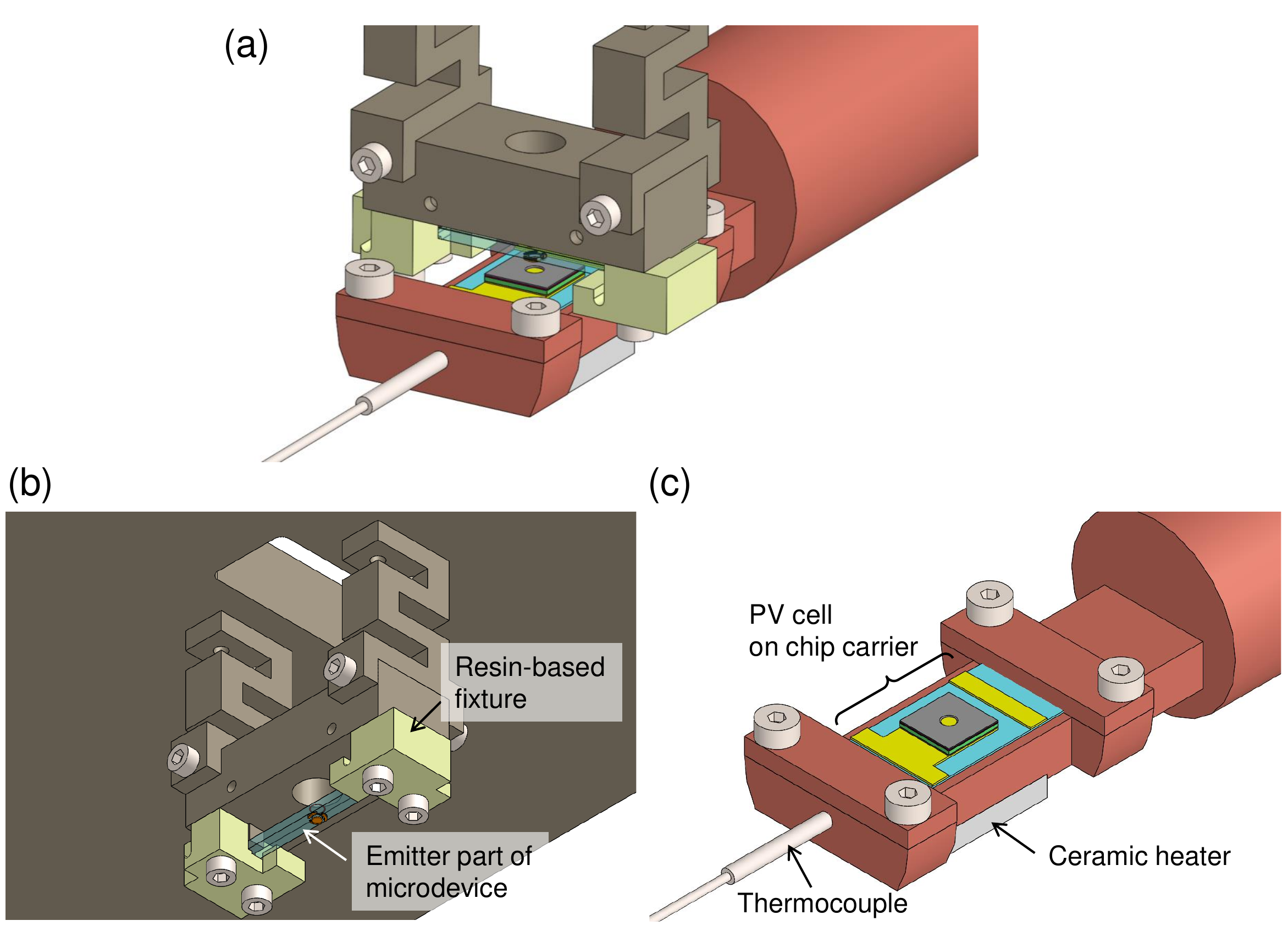}
\caption{\label{FigS5}Supplementary Figure 5: Design illustrations of emitter part of microdevice and PV cell mounted on positioner. (a) Alignment of emitter part of microdevice and PV cell. (b) Emitter part of microdevice mounted on emitter stage. (c) PV cell mounted on receiver stage.}
\end{figure}

\clearpage
\section*{Supplementary Note 3. Emitter temperature estimation and validation}

The temperature of the emitter is determined by finite-element-based transient thermal analysis through ANSYS and using the real data of the heater input power. To repeatedly and consistently raise the emitter temperature, input power is feedback controlled to apply 1.6 W to the Si heater. The consistent change of the electrical resistance during 27 heating repetitions every 6 minutes ensures the stability of the Si heater at high temperatures (see Supplementary Fig.\ \ref{FigS6}a). Supplementary Fig.\ \ref{FigS6}b shows the average of the heater input power and the corresponding resistance change for 27 times measurements. A maximum time variation of 0.11 s is detected during repetitions and denoted as $x$-axis error bars. Transient thermal analysis is conducted using the structural information of the emitter part of the microdevice shown in Supplementary Fig.\ \ref{FigS5}b and the heater input power data described in Supplementary Fig.\ \ref{FigS6}b. The temperature-dependent thermal properties (density, specific heat, and thermal conductivity) of the quartz substrate are obtained from ref.\ \cite{sergeev1982thermophysical}. For the high-temperature-compatible conductive graphite adhesive (931C, Cotronics Corp.) used to connect electrical wires to the heater, the thermal conductivity is obtained from the material's datasheet \cite{931C2006characteristics}, and the density and specific heat are taken from ref.\ \cite{graphite2015characteristics}. We use the temperature-independent thermal properties for the emitter supporter (Inconel alloy 600) \cite{inconel2008characteristics} and the heater (Si) \cite{incropera1996fundamentals} because the temperature of the emitter supporter is maintained around room temperature and the thickness of the heater is negligibly small, i.e., only 800 nm.

In Supplementary Fig.\ \ref{FigS7}a, we can see the temperature of the emitter as a function of the heater input power. In the inset, the uncertainty of the emitter temperature is displayed as a colored band. The width of this band is determined by combining each uncertainty component (i.e., $u=\sqrt{\sum u_i^2}$ where $u_i$ is an uncertainty component). The first uncertainty factor is the heating input time variation, denoted as $x$-axis error bars in Supplementary Fig.\ \ref{FigS6}b. Through the transient thermal simulation, we check that a 0.11 s time variation can cause the emitter temperature uncertainty of about 1.2 K. The second factor is the volume deviation of the graphite adhesive. If there is a 20$\%$ volume deviation, it leads to about 5.6 K emitter temperature uncertainty for the same input conditions. In Supplementary Fig.\ \ref{FigS7}b, the photocurrent generation simulated using Eq.\ (3) of the manuscript is shown as a function of the emitter temperature at the vacuum gap distance of 2.5 $\mu$m. The confidence interval of photocurrent, considering the uncertainty of the emitter temperature, is expressed as a colored band. Since the five photocurrent measurements at corresponding emitter temperatures, which are estimated by transient thermal analysis (i.e., red symbols in Supplementary Fig.\ \ref{FigS7}b), agree well with the theoretical simulation within the colored band, we were able to simultaneously validate the emitter temperature simulation and the photocurrent calculation model. When the emitter temperature is 786 K, considering the heating input time variation and volume deviation of the graphite adhesive, the simulated emitter temperature can be varied by 6 K higher and 4 K lower. This analysis of the emitter temperature is vacuum-gap independent because the amount of near-field radiation is relatively marginal compared to the heating input power. Even when the vacuum gap distance is 200 nm, the near-field radiation between the emitter and the PV cell is only about 0.01 W, which is around 0.6$\%$ of the heating input power of 1.6 W. The emitter temperature deviation is reflected as colored bands for the photocurrent and electrical power output in Figs.\ 3b and c of the manuscript.

\clearpage
\begin{figure}[h]
\centering\includegraphics[width=17cm]{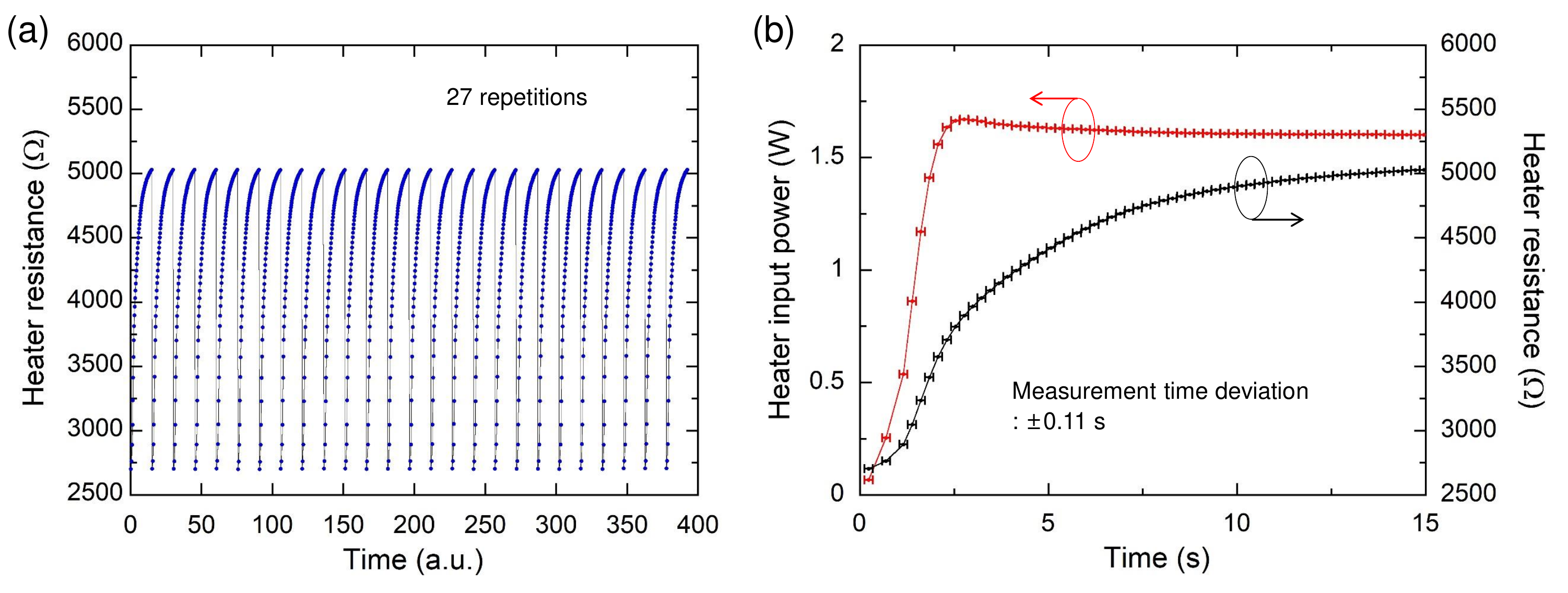}
\caption{\label{FigS6}Supplementary Figure 6: (a) Heater resistance during 27 heating repetitions every 6 minutes. (b) Heater resistance variation with respect to input power.}
\end{figure}

\begin{figure}[h]
\centering\includegraphics[width=17cm]{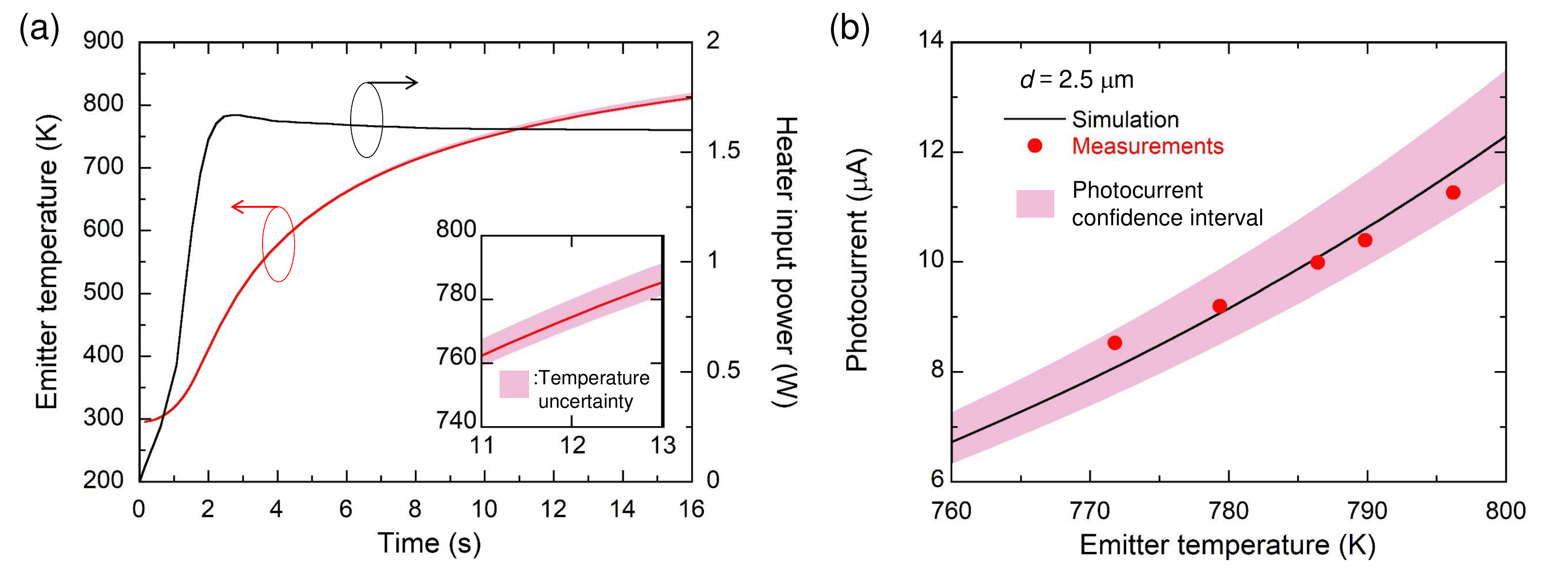}
\caption{\label{FigS7}Supplementary Figure 7: (a) Transient thermal simulation for estimation of emitter temperature. (b) Photocurrent generation with respect to emitter temperature. Transient thermal simulation and photocurrent calculation model are validated by comparing simulation results to measurement results.}
\end{figure}

\clearpage
\section*{Supplementary Note 4. PV cell temperature estimation and current-voltage characterization}

\begin{figure}[h]
\centering\includegraphics[width=17cm]{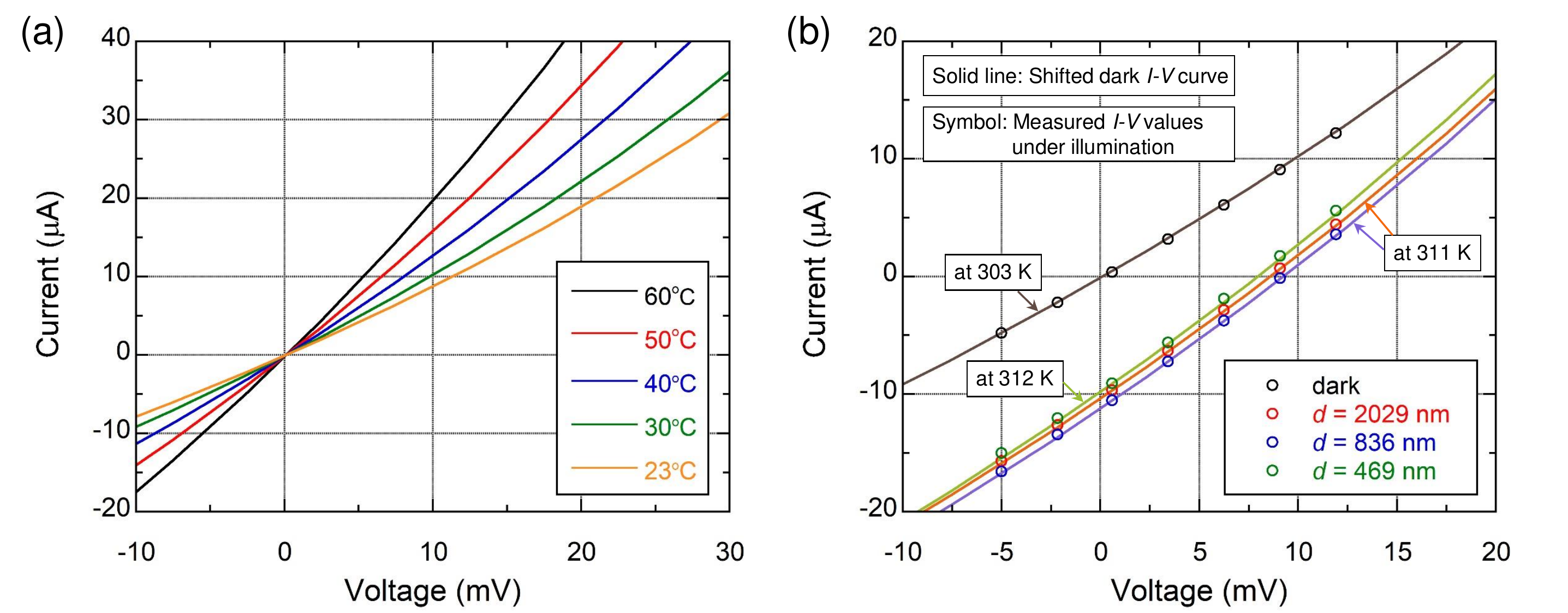}
\caption{\label{FigS8}Supplementary Figure 8: (a) Current-voltage ($I$-$V$) characteristics of Au/$n$-GaSb Schottky PV cell at five different temperatures. (b) Comparison between dark $I$-$V$ curves and measured $I$-$V$ values under illumination conditions. The temperature of the PV cell can be estimated from the slope of the $I$-$V$ characteristics.}
\end{figure}

As is widely known, the diode current at a certain voltage can be used as a temperature sensor. Using this idea, to measure the PV cell temperature, we collected the $I$-$V$ characteristics of the Au/$n$-GaSb Schottky PV cell according to the temperature. The PV cell is heated by a ceramic heater attached to the bottom of the receiver stage, and the corresponding temperature is measured using a thermocouple inserted in the receiver stage. Supplementary Fig.\ \ref{FigS8}a shows several temperature-dependent $I$-$V$ curves among ones collected every 2 minutes while the PV cell is cooled down to room temperature after being heated to 339 K under dark condition. These $I$-$V$ curves are clearly differentiated for each temperature. As in Supplementary Fig.\ \ref{FigS8}b, the temperature of the PV cell can be estimated by shifting the dark $I$-$V$ curve as much as the short-circuit current (i.e., solid lines) and comparing them with the $I$-$V$ values measured under illumination condition (i.e., symbols). At 303 K temperature, the repeatability of the dark $I$-$V$ curve measurement is confirmed by matching the line and symbols. At illumination conditions, when the vacuum gap distance is 2029 nm and 836 nm, the PV cell temperature rises by 8 K from the 303 K. When the vacuum gap becomes 469 nm, the temperature of the PV cell can further rise due to the more contribution of the evanescent mode, but the increasing amount is only 1 K. Therefore, we neglected the change of optical properties resulted from the vacuum-gap-dependent near-field radiation.

At the temperature above 240 K, the $J$-$V$ characteristics of the Au/$n$-GaSb Schottky diode can be expressed using two current mechanisms as follows \cite{jang2021analysis}:
\begin{equation} \label{eq.Jtot}
\begin{split}
J_{tot}(V) & = J_\text{TE}(V) + J_\text{SRH}(V) \\
& = J_{\text{TE}(0)}\Big\lbrace\text{exp}\Big[\frac{e(V-IR_s)}{nk_BT}\Big]-1\Big\rbrace + J_{\text{SRH}(0)}\Big\lbrace\text{exp}\Big[\frac{e(V-IR_s)}{2k_BT}\Big]-1\Big\rbrace
\end{split}
\end{equation}
where $J_{\text{TE}(0)}$ and $J_{\text{SRH}(0)}$ are saturation current density of thermionic emission and SRH recombination, respectively. $e$ is an electron charge, $R_s$ is a series resistance, and $n$ is an ideality factor. The saturation current density of thermionic emission current can be written by \cite{sze2021physics}:
\begin{equation}
J_{\text{TE}(0)} = A^* T^2 \text{exp}\Big(-\frac{e\phi_b}{k_BT}\Big)
\end{equation}
where $A^* = 5.16 \times 10^4$ Am$^{-2}$K$^{-2}$ is a Richardson constant \cite{liu2004improved} and $\phi_b$ is a Schottky barrier height. The saturation current density of SRH recombination is expressed as in the following equation \cite{sze2021physics}:
\begin{equation}
J_{\text{SRH}(0)} = \frac{eW_Dn_i}{2\tau_\text{SRH}}
\end{equation}
where $W_D$ is the width of the depletion region, $n_i$ is an intrinsic carrier concentration, and $\tau_\text{SRH}$ is a SRH recombination lifetime. The parameters that constitute the total current density can be determined through the multi-current fitting method \cite{donoval1991analysis,jang2021analysis} and fitting results are described in Supplementary Figs.\ \ref{FigS9}a and b at 303 and 311 K, respectively. The total current density at the corresponding PV cell temperature given in Eq.\ (\ref{eq.Jtot}) can be used as a dark current density to determine the $I$-$V$ characteristics under the illumination condition.

\begin{figure}[h]
\centering\includegraphics[width=17cm]{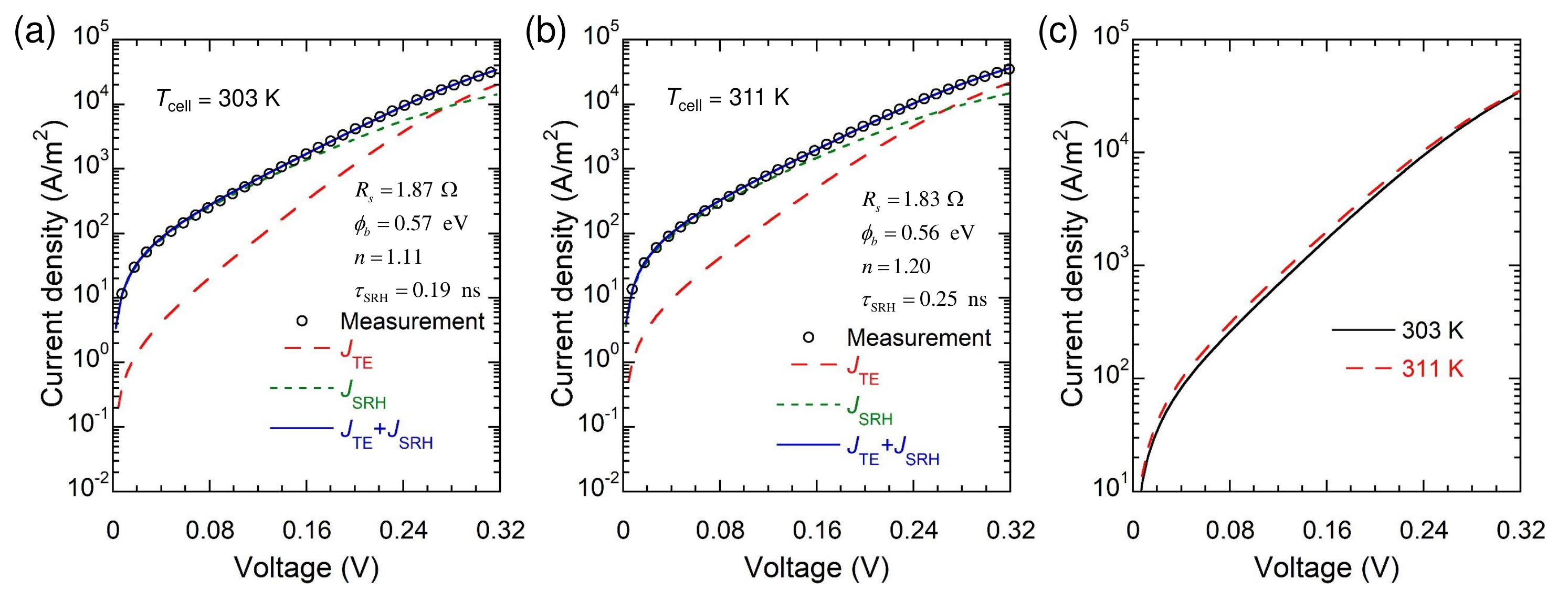}
\caption{\label{FigS9}Supplementary Figure 9: (a,b) Current density and voltage ($J$-$V$) characteristics of Au/$n$-GaSb Schottky PV cell at 303 and 311 K. (c) Comparison of $J$-$V$ curves at 303 and 311 K.}
\end{figure}

\clearpage
\section*{Supplementary Note 5. Data processing and uncertainty analysis}

\begin{figure}[h]
\centering\includegraphics[width=17cm]{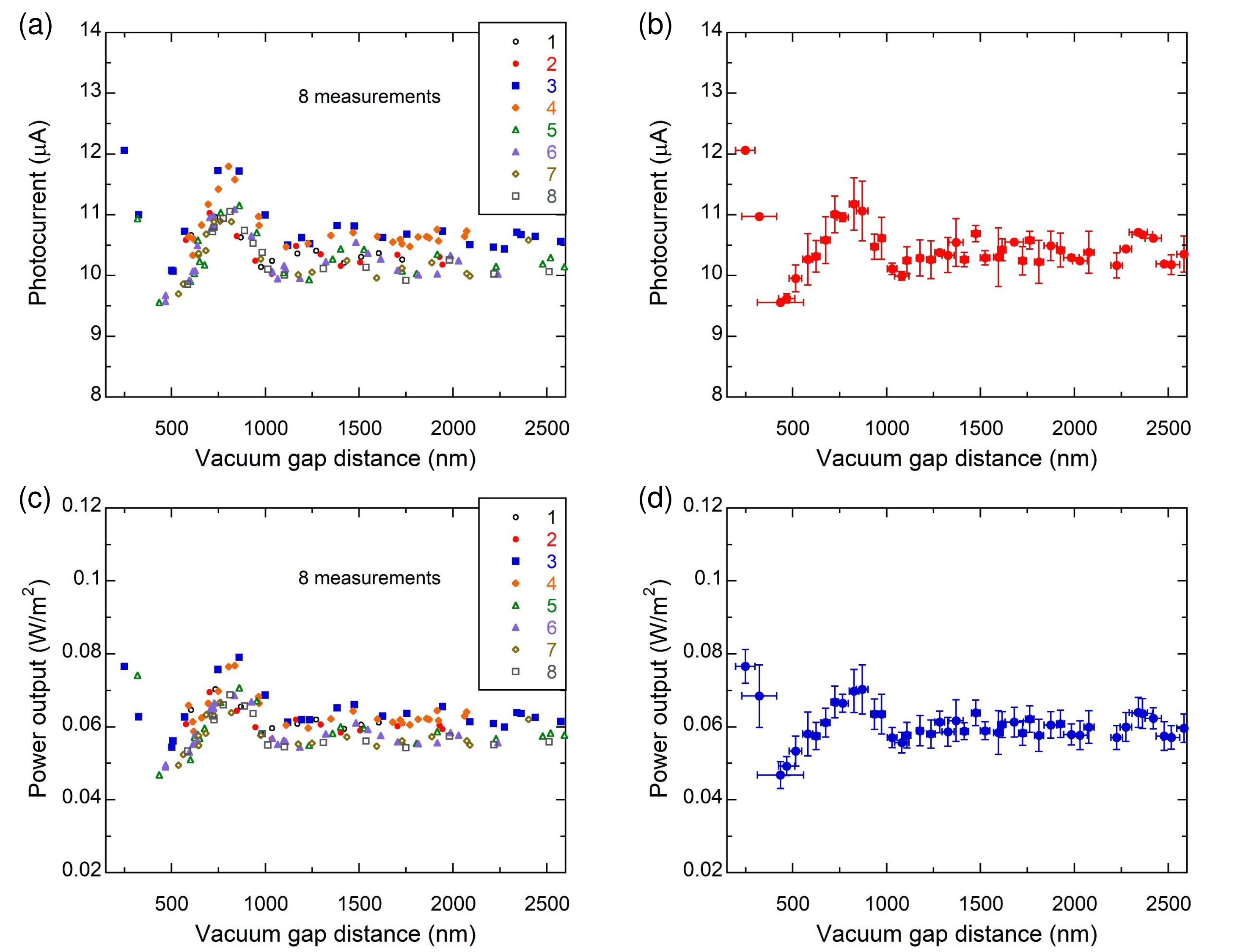}
\caption{\label{FigS10}Supplementary Figure 10: (a) Eight photocurrent measurements with respect to vacuum gap distance. (b) Processed photocurrent data. (c) Eight power output measurements with respect to vacuum gap width. (d) Processed power output data.}
\end{figure}

To process the 8-times repeated measurements of the photocurrent (or electrical power output) according to the vacuum gap distance, first, the measured data is separated into 50-nm-vacuum-gap intervals. Then, the arithmetic averaged vacuum gap distance and photocurrent (or electrical power output) within each interval is designated as a processed value. The length of $x$-axis and $y$-axis error bars of each processed value represent the total uncertainty of the vacuum gap and the photocurrent (or electrical power output), respectively. These values are calculated using the following equation:

\begin{equation} \label{eq.uncer}
u_{tot}=\sqrt{u_a^2+\frac{1}{N^2}\sum_{k=1}^{N}\left[u_b(x_k)\right]^2}
\end{equation}

where $u_a$ is the standard deviation of the data and $N$ is the number of data in each interval. $u_b(x_k)$ is the uncertainty of every single raw datum, $x_k$. $x$ can be the vacuum gap distance $d$, photocurrent $I_{ph}$, or electrical power output $P_E$. For the vacuum gap distance, $u_b(d_k)=\sqrt{\left[u_1(d_k)\right]^2+\left[u_2(d_k)\right]^2}$, where $u_1(d_k)$ denotes the maximum vacuum gap variation that can occur during $I$-$V$ measurement. $u_2(d_k)=\frac{\partial d_k}{\partial C_k}u(C_k)$ represents the measurement uncertainty of the vacuum gap distance where $C$ is the measured capacitance and $u(C_k)$ is the measurement accuracy of the instrument (E4980AL, Keysight). For the photocurrent, $u_b(I_{ph,k})$ is also determined by the measurement accuracy of the instrument (2400 Sourcemeter, Keithley). Since the electrical power output is calculated as the product of the current and the voltage, we can obtain the uncertainty of a single power output datum as following: $u_b(P_{E,k})=\sqrt{\left[\frac{\partial P_{E,k}}{\partial V_k}u(V_k)\right]^2+\left[\frac{\partial P_{E,k}}{\partial I_k}u(I_k)\right]^2}$, where $u(V_k)$ and $u(I_k)$ are the measurement accuracy of the instrument (2400 Sourcemeter, Keithley). Supplementary Figs.\ \ref{FigS10}a and c show the original data of the photocurrent and the electrical power output for the 8 repeated experiments. After data processing using the arithmetic average and Eq.\ (\ref{eq.uncer}), we can describe the processed data as in Supplementary Figs.\ \ref{FigS10}b and d, with error bars for the total uncertainty.

\clearpage
\section*{Supplementary Note 6. Spectral radiative heat flux and photocurrent}

As explained in the methods section of the manuscript, the above-bandgap radiation absorbed by the PV cell can be classified into two parts because the area size of the doped-Si emitter and that of the active area of the PV cell are different. One is the vacuum-gap-dependent radiation between the doped-Si emitter and the PV cell; the other is the far-field radiation between the SiO$_2$ and the PV cell. The spectral radiative heat flux including both parts is shown in Supplementary Fig.\ \ref{FigS11}a for 900 and 550 nm vacuum gap distances. The above-bandgap radiative heat flux of the 900 nm gap is larger than that of the 550 nm gap. On the other hand, the total heat flux containing the sub-bandgap region is larger for the vacuum gap of 550 nm. Supplementary Fig.\ \ref{FigS11}b shows the spectral photocurrent for the two vacuum gap distances. Since the PV cell serves as a bandpass filter, it seems that we can measure the quasi-monochromatic radiation corresponding to the bandgap wavelength.

\begin{figure}[h]
\centering\includegraphics[width=17cm]{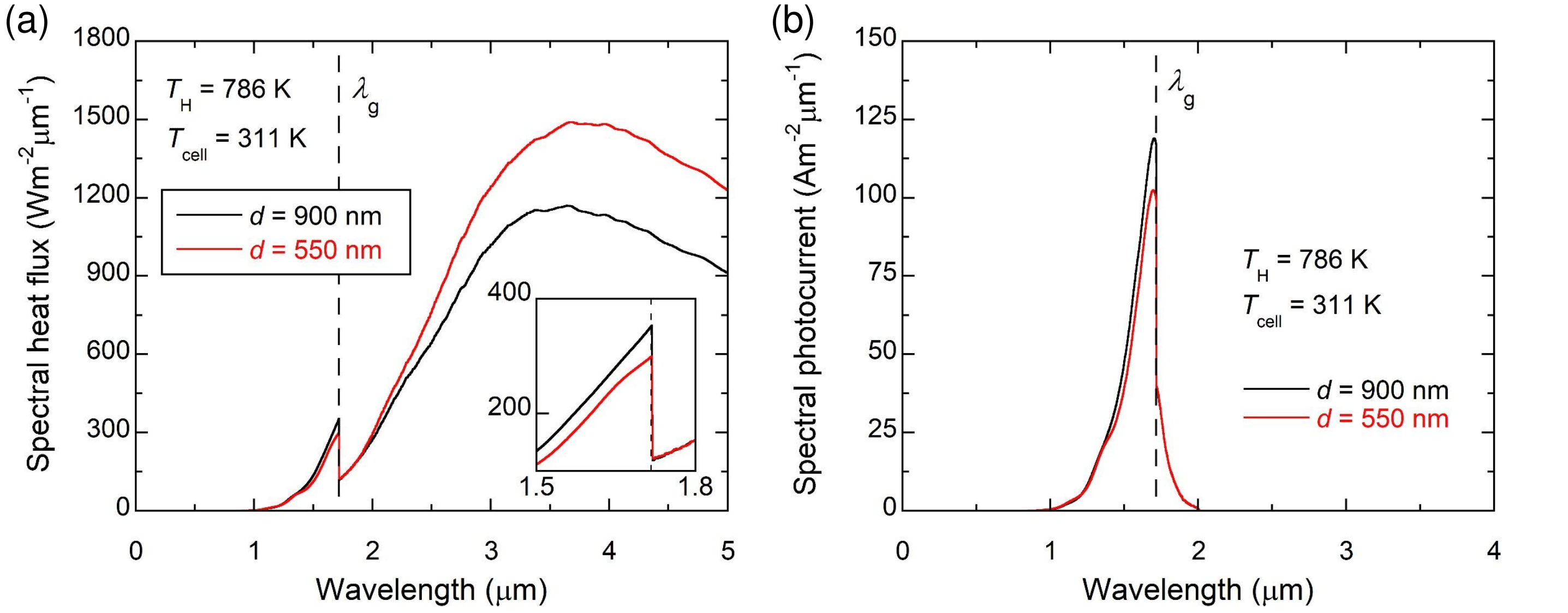}
\caption{\label{FigS11}Supplementary Figure 11: (a) Spectral radiative heat flux absorbed in PV cell. (b) Spectral photocurrent density generated in PV cell.}
\end{figure}

\begin{figure}[h]
\centering\includegraphics[width=9cm]{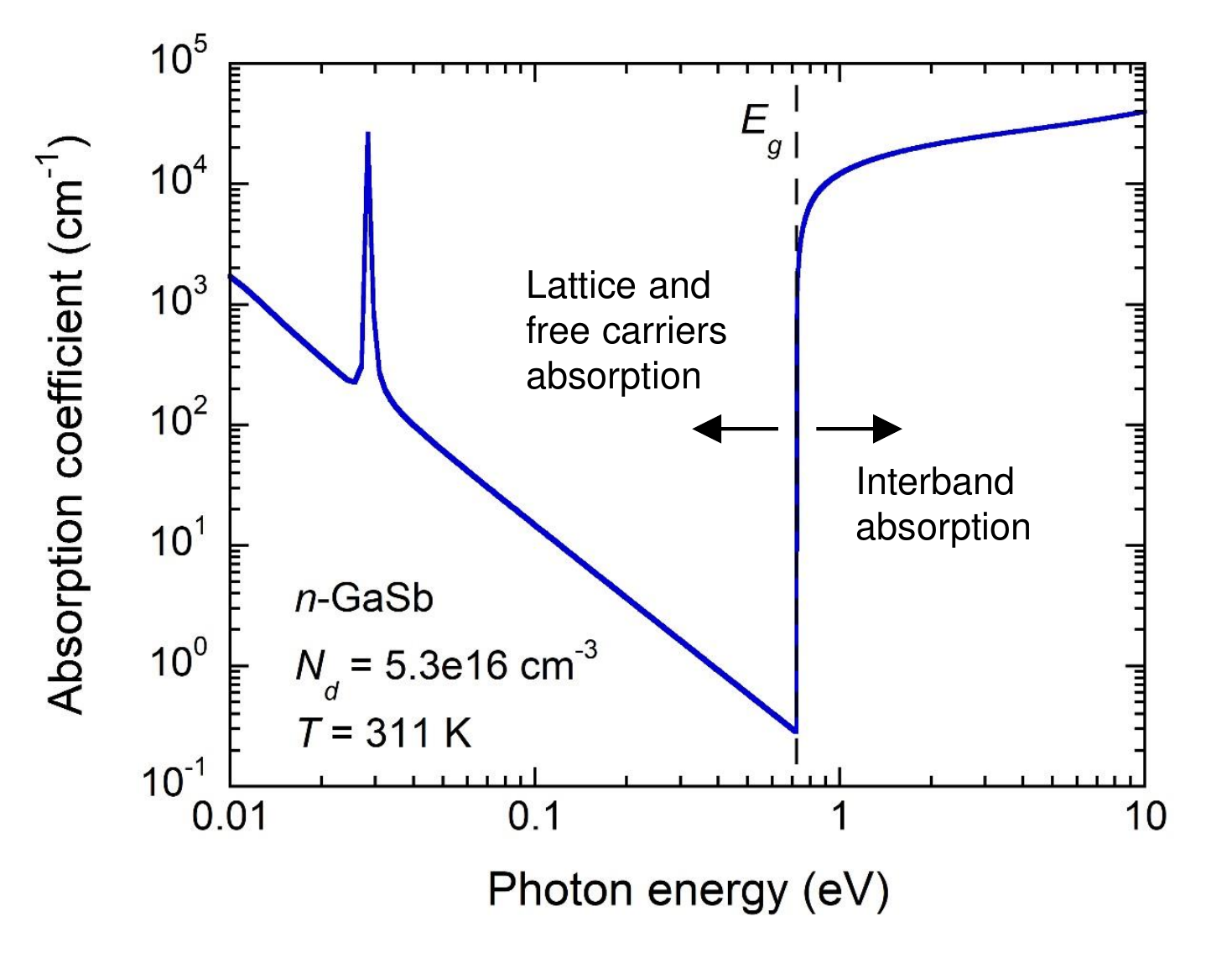}
\caption{\label{FigS12}Supplementary Figure 12: Absorption coefficient of $n$-GaSb when doping concentration is $5.3\times10^{16}$ cm$^{-3}$ and temperature is 311 K.}
\end{figure}

\clearpage
\section*{Supplementary Note 7. Quantum efficiency of Au/$n$-GaSb Schottky PV cell}

To measure the quantum efficiency of the Schottky PV cell, an experimental apparatus shown in Supplementary Fig.\ \ref{FigS13}a consisting of a solar simulator (94011A, Newport) and two filter wheels (FW102C, Thorlabs) is used. As shown in the schematic illustration in Supplementary Fig.\ \ref{FigS13}b, the light emitted from the solar simulator passes through two optical filters [i.e., one long-pass filter (86-073, Edmund Optics) and one band-pass filter] and reaches the Ge photodetector (71SI00422, Newport) or the Schottky PV cell placed on the receiver stage in the vacuum chamber. The Ge photodetector can measure light intensity in a wavelength range from 800 to 1800 nm. Eight band-pass filters (85-897, 85-898, 85-899, 85-900, 85-901, 85-902, 85-903, and 87-872, Edmund Optics), with a 50 nm full width at half maximum (FWHM), are used; they cover the wavelength range of 1250-1600 nm, enabling measurement of the quantum efficiency of the Schottky PV cell in that range. First, we place a Ge photodetector and measure the light intensity. Then, the photocurrent generated by the Schottky PV cell is measured at the same location. Therefore, we can measure the external quantum efficiency (EQE) of a specific wavelength using the following equation:
\begin{equation}
\eta_\text{ext}(\lambda)=I_{ph,\lambda}\cdot\frac{hc_0}{e\lambda}\cdot\frac{A_\text{pd}}{Q_\lambda A_2}
\end{equation}
where $I_{ph,\lambda}$ is the photocurrent generated in the Schottky PV cell, $h$ is Planck constant, $c_0$ is the speed of light in vacuum, $e$ is the electron charge, $A_\text{pd}$ is the area of the Ge photodetector, $Q_\lambda$ is the light intensity, and $A_2$ is the area of the Schottky PV cell. To verify the experimental apparatus, we compare the EQE measured by the Korea Institute of Energy Research (KIER) and that measured by the filter-wheel-based apparatus. As shown in Supplementary Fig.\ \ref{FigS13}c, the values match well, with an error smaller than 1$\%$. Then, the EQE of the Schottky PV cell used in the NF-TPV experiment is measured using a filter-wheel-based apparatus; those values are indicated as red symbols in Supplementary Fig.\ \ref{FigS13}d. The EQE values are compared with the data measured by KIER shown in Supplementary Fig.\ \ref{FigS13}c and the full spectrum EQE are fitted to the red symbols. We are able to find the internal quantum efficiency (IQE) by dividing the EQE into the spectral absorptivity of the Schottky PV cell, obtained using the Fresnel coefficient \cite{zhang2007nano}. According to the Varshni equation \cite{gonzalez2006modeling,varshni1967temperature}, the bandgap energy hardly changes when the GaSb temperature increases from 303 K to 311 K (i.e., $\lambda_g$ changes from 1.71 $\mu$m to 1.72 $\mu$m). Therefore, the quantum efficiency obtained here can be safely used for the theoretical analysis.

\begin{figure}[h]
\centering\includegraphics[width=17cm]{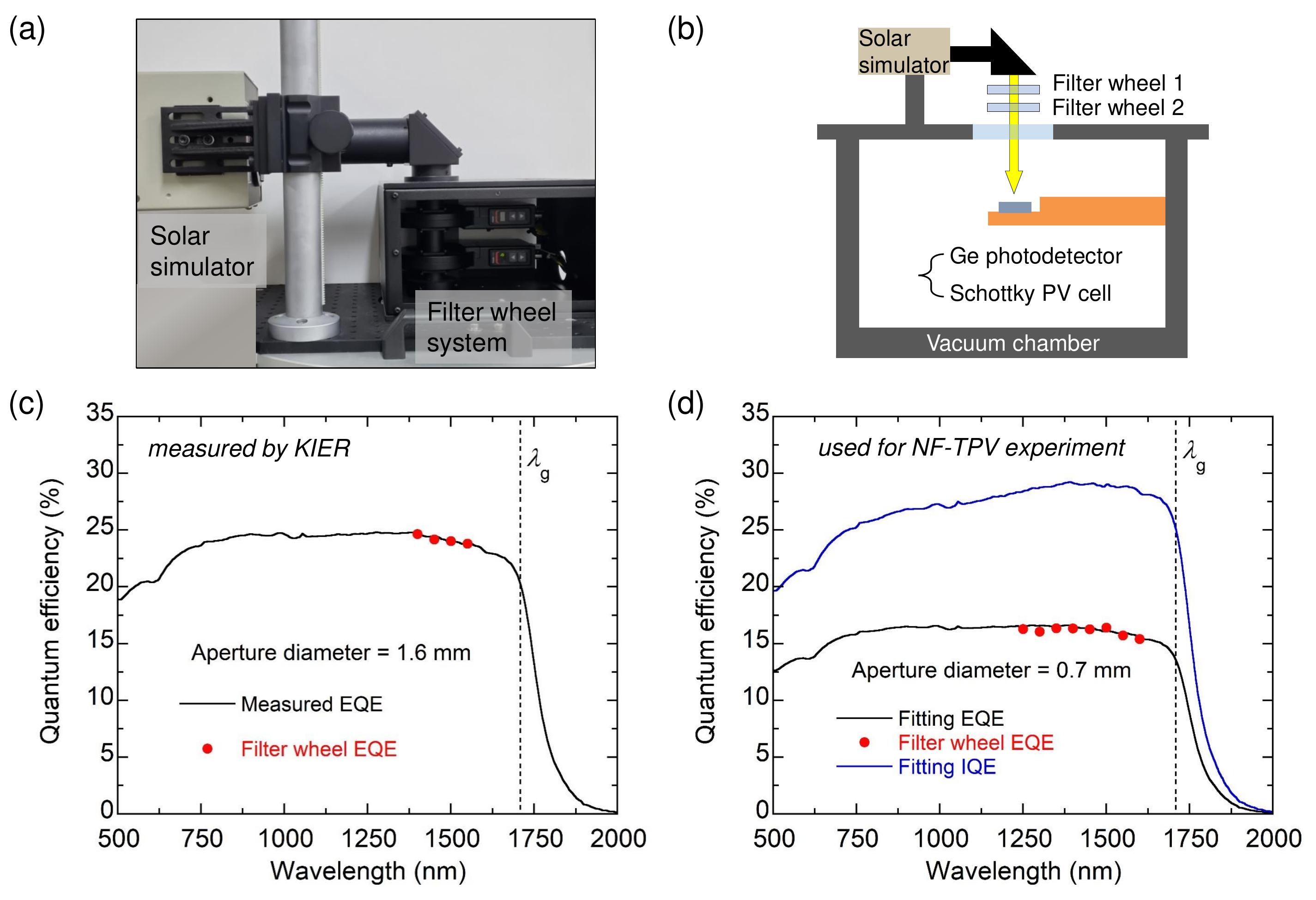}
\caption{\label{FigS13}Supplementary Figure 13: (a) Experimental apparatus for quantum efficiency measurement. (b) Schematic of quantum efficiency measurement apparatus. (c) Quantum efficiency of 1.6-mm-diameter Au/$n$-GaSb Schottky PV cell. (d) Quantum efficiency of 0.7-mm-diameter Au/$n$-GaSb Schottky PV cell.}
\end{figure}

\clearpage

\providecommand{\noopsort}[1]{}\providecommand{\singleletter}[1]{#1}%